\input harvmac

\input epsf


\overfullrule=0pt

\def\frac#1#2{{#1\over #2}}

\def\IP{{\bf P}}

\def\bfone{\relax{\rm 1\kern-.35em 1}}
\def\inbar{\vrule height1.5ex width.4pt depth0pt}
\def\IC{\relax\,\hbox{$\inbar\kern-.3em{\mss C}$}}
\def\ID{\relax{\rm I\kern-.18em D}}
\def\IF{\relax{\rm I\kern-.18em F}}
\def\IH{\relax{\rm I\kern-.18em H}}
\def\II{\relax{\rm I\kern-.17em I}}
\def\IN{\relax{\rm I\kern-.18em N}}
\def\IQ{\relax\,\hbox{$\inbar\kern-.3em{\rm Q}$}}
\def\us#1{\underline{#1}}
\def\IR{\relax{\rm I\kern-.18em R}}
\font\cmss=cmss10 \font\cmsss=cmss10 at 7pt
\def\ZZ{\relax\ifmmode\mathchoice
{\hbox{\cmss Z\kern-.4em Z}}{\hbox{\cmss Z\kern-.4em Z}}
{\lower.9pt\hbox{\cmsss Z\kern-.4em Z}}
{\lower1.2pt\hbox{\cmsss Z\kern-.4em Z}}\else{\cmss Z\kern-.4em
Z}\fi}
\def\nup#1({Nucl.\ Phys.\ $\us {B#1}$\ (}
\def\plt#1({Phys.\ Lett.\ $\us  {B#1}$\ (}
\def\cmp#1({Comm.\ Math.\ Phys.\ $\us  {#1}$\ (}
\def\prp#1({Phys.\ Rep.\ $\us  {#1}$\ (}
\def\prl#1({Phys.\ Rev.\ Lett.\ $\us  {#1}$\ (}
\def\prv#1({Phys.\ Rev.\ $\us  {#1}$\ (}
\def\mpl#1({Mod.\ Phys.\ Let.\ $\us  {A#1}$\ (}
\def\ijmp#1({Int.\ J.\ Mod.\ Phys.\ $\us{A#1}$\ (}
\def\tit#1|{{\it #1},\ }

\def\Coe#1.#2.{{#1\over #2}}

\def\coe#1.#2.{\relax{\textstyle {#1 \over #2}}\displaystyle}
\def\half{{1 \over 2}}

\def\dd{{\rm d}}

\lref\ghovaf{D. Ghoshal and C. Vafa,
``c=1 String as the Topological Theory of the Conifold,''
Nucl. Phys. {\bf B453} (1995) 121.}

\lref\quot{I. Ciocan-Fontanine and M. M. Kapranov, ``Derived Quot Schemes,''
math.AG/9905174.}

\lref\kthy{E. Witten, ``D-Branes and K-Theory,'' hep-th/9810188.}

\lref\harveymoore{}

\lref\marinomoore{M Marino and G. Moore,
``Counting Higher Genus Curves in a Calabi-Yau Manifold,''
Nucl. Phys. {\bf B543} (1999) 592-614; hep-th/9808131.}

\lref\hst{S. Hosono, M.-H. Saito, and A. Takahashi, ``Holomorphic
Anomaly Equation and BPS State Counting of Rational Elliptic
Surface,'' hep-th/9901151.}

\lref\swrefs{
W. Lerche, 
``Introduction to Seiberg-Witten Theory and its Stringy Origin,'' 
Nucl. Phys. Proc. Suppl. {\bf 55B} (1997) 83-117, hep-th/9611190; 
A. Klemm, ``On the Geometry behind N=2 Supersymmetric Effective Actions in 
Four-Dimensions'', Proceedings of the Trieste Summer School on 
High Energy Physics and Cosmology {\bf 96}, World Scientic, Singapore (1997) 
120-242, 
hep-th/9705131; P. Mayr, ``Geometric Construction of N=2 Gauge Theories,'' 
Fortsch. Phys. {\bf 47} (1999) 39-63, hep-th/9807096.}

\lref\kappriv{M. M. Kapranov, private communication.}

\lref\aspdon{P. S. Aspinwall and R. Y. Donagi, ``The Heterotic
String, the Tangent Bundle, and Derived Categories,''
Adv. Theor. Math. Phys. {\bf 2} (1998) 1041-1074;
hep-th/9806094.}

\lref\aspmor{P. Aspinwall and D. Morrison, ``Topological Field
Theory and Rational Curves,'' Commun. Math. Phys. {\bf 51} (1993)
245-262.}

\lref\manin{Yu. I. Manin, ``Generating Functions in Algebraic Geometry and Sums 
over Trees'' in ``The moduli space of curves''
(Texel Island, 1994), Progr. Math. {\bf 129}, 
Birkh\"auser, Boston (1995) 401--417; alg-geom/9407005.}

\lref\morvaf{D. R. Morrison and C. Vafa, ``Compactifications
of F-Theory on Calabi-Yau threefolds -- II,''
Nucl. Phys. {\bf B476} (1996) 437-469; hep-th/9603161.}

\lref\minmoor{R. Minasian and G. Moore, ``K-Theory and
Ramond-Ramond Charge,'' hep-th/9710230.}

\lref\nek{N. Nekrasov,  ``In the Woods of M-Theory,'' hep-th/9810168.}

\lref\vafaoog{H. Ooguri and C. Vafa, ``Summing up Dirichlet
Instantons,''
Phys. Rev. Lett. {\bf 77} (1996) 3296-3298.}

\lref\cfkmII{P. Candelas, A. Font, S. Katz, and D. R. Morrison
    ``Mirror Symmetry for Two Parameter Models -- II,''
      Nucl. Phys. {\bf B429} (1994) 626-674.}

\lref\vertalgs{See, e.g., H. Nakajima, ``Lectures on Hilbert Schemes
of Points on Surfaces,''
preprint http://kusm.kyoto-u.ac.jp/~nakajima/TeX.html;
I. Grojnowski, ``Instantons and Affine Algebras I:  The Hilbert
Scheme and Vertex Operators,'' Math. Res. Lett. {\bf 3}, no. 2 (1996)
275-291; and V. Baranovsky, ``Moduli of Sheaves on Surfaces and Action
of the Oscillator Algebra,'' math.AG/9811092.}

\lref\vafaslag{C. Vafa, ``Extending Mirror Conjecture to Calabi-Yau
with Bundles,'' hep-th/9804131.}

\lref\cdfkmI{P. Candelas, X. de la Ossa, A. Font, S. Katz, and D. R.
Morrison
    ``Mirror Symmetry for Two Parameter Models -- I,''
      Nucl.Phys. {\bf B416} (1994) 481-538.}

\lref\grabp{T. Graber, private communication.}

\lref\katzcox{D. A. Cox, and S. Katz, ``{\it Mirror Symmetry and
Algebraic
Geometry},'' AMS, Providence, RI USA, 1999, to appear.}

\lref\vain{I. Vainsencher, ``Enumeration of n-fold Tangent
Hyperplanes to a Surface,'' J. Algebraic Geom. {\bf 4},
1995, 503-526, math.AG/9312012.}

\lref\fultoni{W. Fulton, ``{\it Intersection Theory},'' Second edition,
Springer-Verlag (Berlin) 1998.}

\lref\faber{C. Faber, ``Algorithm for Computing Intersection Numbers
on Moduli Spaces of Curves, with an Application to the Class of
the Locus of the Jacobians,'' math.AG/9706006.}

\lref\fabpan{C. Faber and R. Pandharipande, ``Hodge Integrals and Gromov-Witten
Theory,'' math.AG/9810173.}

\lref\penglu{P. Lu, ``Special Lagrangian Tori on a
Borcea-Voisin Threefold,'' math.DG/9902063.}

\lref\vafagop{C. Vafa and R. Gopakumar, ``M-theory and
Topological Strings-I \& II,'' hep-th/9809187, hep-th/9812127.}

\lref\litian{J. Li and G. Tian, ``Virtual Moduli Cycles
and Gromov-Witten Invariants of Algebraic Varieties,''
J. Amer. Math. Soc. {\bf 11}, no. 1 (1998) 119-174.}

\lref\pandh{R. Pandharipande, ``Hodge Integrals and
Degenerate Contributions,'' math.AG/9811140.}

\lref\klemmhg{A. Klemm, ``Mirror Symmetry at Higher Loops: A Precision
Test for M-theory/IIA Duality,'' to appear.}

\lref\behfan{K. Behrend and B. Fantechi, ``The Intrinsic
Normal Cone,'' math.AG/9601010.}

\lref\locvir{T. Graber and R. Pandharipande, ``Localization
of Virtual Classes,'' math.AG/9708001.}

\lref\kontloc{M. Kontsevich, ``Enumeration of Rational Curves
via Torus Actions,'' in {\sl The Moduli Space of Curves,}
Dijkgraaf et al eds., Progress in Mathematics {\bf 129},
Birkh\"auser (Boston) 1995.}

\lref\gasd{ C. Vafa, ``Gas of D-Branes and Hagedorn Density of BPS
States,''
Nucl. Phys. {\bfB463} (1996) 415-419; hep-th/9511088.}

\lref\bsv{ M. Bershadsky, V. Sadov, and C. Vafa, ``D-Branes and
Topological
Field Theories,'' Nucl. Phys. {\bf B463} (1996) 420-434;
hep-th/9511222.}

\lref\yz{S.-T. Yau and E. Zaslow, ``BPS States, String Duality, and
Nodal
Curves on K3,'' Nucl. Phys. {\bf B471} (1996) 503-512; hep-th/9512121.}

\lref\yztwo{ S.-T. Yau
and E. Zaslow, ``BPS States as Symplectic Invariants from String
Theory,''  in {\sl Geometry and Physics,} Proceedings of the
Special Session on Geometry and Physics, Aarhus, Denmark, 1996.}

\lref\cdgp{ P. Candelas, X. C. De La Ossa, P. Green, and L. Parkes,
``A Pair of Calabi-Yau Manifolds as an Exactly Soluble Superconformal
Theory,'' Nucl. Phys. {\bf B359} (1991) 21.}

\lref\mcl{R. McLean, ``Deformations of Calibrated Submanifolds,"
Duke preprint 96-01:  www.math.duke.edu/preprints/1996.html.}

\lref\hl{F. R. Harvey and H. B. Lawson, ``Calibrated Geometries,''
Acta Math. {\bf 148} (1982) 47;
F. R. Harvey, {\sl Spinors and Calibrations,} Academic Press,
New York, 1990.}

\lref\ooguri{ K. Becker, M. Becker, D. R. Morrison, H. Ooguri,
Y. Oz, and Z. Yin, ``Supersymmetric Cycles in Exceptional Holonomy
Manifolds and Calabi-Yau Fourfolds,'' Nucl. Phys. {\bf B480} (1996)
225-238.}

\lref\vafmir{ C. Vafa, ``Extending Mirror Conjecture to Calabi-Yau
with Bundles,'' hep-th/9804131.}

\lref\zaslow{ E. Zaslow, ``Solitons and Helices:  The Search for
a Math-Physics Bridge,'' Commun. Math. Phys. {\bf 175} (1996) 337-375.}

\lref\kach{ S. Kachru, A. Klemm, W. Lerche, P. Mayr, C. Vafa,
``Nonperturbative Results on the Point Particle Limit of N=2 Heterotic
String,''
Nucl. Phys. {\bf B459} (1996) 537.}

\lref\helix{  For several articles on the subject of helices,
see Rudakov, A. N. et al, {\sl Helices and Vector Bundles.}
Seminaire Rudakov, London Mathematical Society Lecture Note
Series {\bf 148,} Cambridge University Press (Cambridge), 1990.}

\lref\kkv{ S. Katz, A. Klemm, and C. Vafa,
``Geometric Engineering of Quantum Field Theories,''
Nucl. Phys. {\bf B497} (1997) 173-195.}

\lref\wittdgrav{E.Witten,
``Two-dimensional Gravity and Intersection Theory on Moduli Space,''
Surveys in Differential Geometry {\bf 1} (1991) 243-310.}

\lref\lly{ B. Lian, K. Liu, and S.-T. Yau, ``Mirror Principle I,''
Asian J. of Math. Vol. {\bf 1} No. 4 (1997) 729-763; math.AG/9712011.}

\lref\llyt{ B. Lian, K. Liu, and S.-T. Yau, ``Mirror Principle II,''
in preparation.}

\lref\givental{ A. Givental, ``A Mirror Theorem for Toric Complete
Intersections,''
{\it Topological Field Theory, Primitive Forms and Related Topics
(Kyoto,
1996)}, Prog. Math. {\bf 160}, 141-175; math.AG/9701016.}

\lref\syz{ A. Strominger,
S.-T. Yau, and E. Zaslow, ``Mirror Symmetry is T-Duality,''
Nuclear Physics {\bf B479} (1996) 243-259; hep-th/9606040.}

\lref\slag{ D. Morrison, ``The Geometry Underlying Mirror Symmetry,''
math.AG/9608006;
M. Gross and P. Wilson, ``Mirror Symmetry via 3-tori for a Class of
Calabi-Yau
Threefolds,'' to appear in Math. Ann., math.AG/9608009;  B. Acharya, ``A

Mirror Pair of Calabi-Yau Fourfolds in Type II String Theory,'' Nucl.
Phys.
{\bf B524} (1998) 283-294, hep-th/9703029;
N. C. Leung and C. Vafa, ``Branes and Toric Geometry,'' Adv. Theor.
Math.
Phys. {\bf 2} (1998) 91-118, hep-th/9711013;
N. Hitchin, ``The Moduli Space of Special Lagrangian Submanifolds,''
math.DG/9711002.}

\lref\kontsevich{ M. Kontsevich, ``Homological Algebra of Mirror
Symmetry,''
Proceedings of the 1994 International Congress of Mathematicians {\bf
I},
Birk\"auser, Z\"urich, 1995, p. 120;
math.AG/9411018.}

\lref\kmvt{ S. Katz, P. Mayr and C. Vafa, ``Mirror Symmetry and
Exact Solution of 4D N=2 Gauge Theories - I,'' Adv. Theor.
Math. Phys. {\bf 1} (1997) 53-114.}

\lref\kkmv{ S. Katz, A. Klemm, and C. Vafa, ``Geometric Engineering of
Quantum Field Theories,'' Nucl. Phys. {\bf B497} (1997) 173-195,
hep-th/9609239;
and A. Klemm, P. Mayr, and C. Vafa, ``BPS States of Exceptional
Non-Critical Strings,''
hep-th/9607139.}

\lref\guzz{ D. Guzzetti, ``Stokes Matrices and Monodromy for the
Quantum Cohomology of Projective Spaces,'' preprint SISSA 87/98/FM.}

\lref\dub{ B. Dubrovin,
{\sl Geometry of 2D Topological Field Theories,}
Lecture Notes in Math {\bf 1620} (1996) 120-348.}

\lref\ttstar{ S. Cecotti and C. Vafa, ``Topological Anti-Topological
Fusion,'' Nucl. Phys. {\bf B367} (1991) 359-461.}

\lref\class{ S. Cecotti and C. Vafa, ``On Classification of
N=2 Supersymmetric Theories,'' Commun. Math. Phys. {\bf 158} (1993)
569-644.}

\lref\dubconj{ ``Geometry and Analytic Theory of Frobenius
Manifolds,'' math.AG/9807034.}

\lref\ad{ P. Aspinwall and R. Dongagi,
``The Heterotic String, the Tangent Bundle,
and Derived Categories,'' hep-th/9806094.}

\lref\candelas{P.\  Candelas, X. C. De La Ossa, P. S. Green, L. Parkes,
``A Pair
of Calabi-Yau Manifolds as an Exactly Soluble Superconformal Theory,''
Nucl.Phys. {\bf B359} (1991) 21-74.}

\lref\batyrev{V.\ Batyrev, ``Dual Polyhedra and Mirror Symmetry for
Calabi-Yau Hypersurfaces
in Toric Varieties,'' J. Algebraic Geom. {\bf 3} (1994) 493-535.}

\lref\batyrevII{V.\ Batyrev, ``Variations of the Mixed Hodge Structure
of Affine Hypersurfaces in Algebraic Tori,''
Duke Math. Jour. {\bf 69}, 2 (1993) 349}

\lref\globalcalabiyau{ S.  Hosono, A.  Klemm, S. Theisen and S.T. Yau,
``Mirror Symmetry, Mirror Map and Applications to Calabi-Yau
Hypersurfaces,'' Commun. Math. Phys. {\bf 167} (1995) 301-350, hep-th/9308122;
P.\  Candelas, A. Font, S. Katz and D. Morrison,
``Mirror Symmetry for Two Parameter Models,''
Nucl.Phys. {\bf B429} (1994) 626-674, hep-th/9403187.}

\lref\cox{D.\ Cox, ``The Homogeneous Coordinate Ring of a Toric
Variety,''
J.\ Alg. Geom {\bf 4} (1995) 17; math.AG/9206011}

\lref\danilov{V. I. Danilov, {\sl The Geometry of Toric Varieties,}
Russian Math. Surveys,{\bf  33} (1978) 97.}

\lref\oda{T.\ Oda,  {\sl Convex Bodies and Algebraic Geometry, An
Introduction to the
Theory of Toric Varieties,} Ergennisse der Mathematik und ihrer
Grenzgebiete, 3. Folge, Bd. {\bf 15},
Springer-Verlag (Berlin) 1988.}

\lref\fulton{W.\ Fulton, {\sl Introduction to Toric Varieties,}
Princeton Univ. Press {\bf 131} (Princeton) 1993}

\lref\batyrevborisov{V.\ Batyrev and L.\ Borisov, ``On Calabi-Yau
Complete Intersections in
Toric Varieties,'' in {\sl Higher-dimensional Complex Varieties,}
(Trento, 1994), 39-65, de Gruyter  (Berlin) 1996.}

\lref\secondaryfan{I. M. Gel'fand, M. Kapranov, A. Zelevinsky, {\sl
Multidimensional Determinants,
Discriminants and Resultants}, Birkh\"auser (Boston) 1994;
L. Billera, J. Filiman and B.\ Sturmfels, ``
Constructions and Complexity of Secondary Polytopes,'' Adv. Math. {\bf
83}
(1990),  155-17.}

\lref\kmv{A. Klemm, P. Mayr, and C. Vafa, ``BPS States of Exceptional
Non-Critical Strings,'' Nucl. Phys. {\bf B}
(Proc. Suppl.) 58 (1997) 177-194; hep-th/9607139.}

\lref\mnw{J. A. Minahan, D. Nemechansky and N. P. Warner, ``Partition
function for the BPS States
of the Non-Critical $E_8$ String,'' Adv. Theor. Math.Phys. {\bf 1}
(1998) 167-183; hep-th/9707149.}

\lref\mnvw{J. A. Minahan, D. Nemechansky, C. Vafa and N. P. Warner,
``$E$-Strings and $N=4$ Topological Yang-Mills Theories,''
Nucl. Phys. {\bf B527} (1998) 581-623.}

\lref\lmw{W. Lerche, P. Mayr, and N. P. Warner,
``Non-Critical Strings, Del Pezzo Singularities
and Seiberg-Witten Curves,'' hep-th/9612085.}

\lref\morrison{D.\ Morrison,
``Where is the large Radius Limit?'' Int. Conf. on Strings 93,
Berkeley; hep-th/9311049}
\lref\griffith{P. Griffiths, ``On the Periods of certain
Rational Integrals,'' Ann. Math. {\bf 90} (1969) 460.}

\lref\klty{A. Klemm, W. Lerche, S. Theisen and S. Yankielowicz,
      ``Simple Singularities and N=2 Supersymmetric Yang-Mills Theory,''

        Phys. Lett. {\bf B344} (1995) 169; A.Klemm, W.Lerche, S.Theisen,

       ``Nonperturbative Effective Actions of N=2 Supersymmetric Gauge
Theories,''
       J.Mod.Phys. {\bf A11} (1996) 1929-1974}

\lref\linearsigmamodel{E. Witten, ``Phases of $N=2$ Theories in Two
Dimensions,'' Nucl. Phys. {\bf B403} (1993) 159.}

\lref\klry{A. Klemm, B. Lian, S.-S. Roan, S.-T. Yau. ``
         Calabi-Yau fourfolds for M- and F-Theory compactifications, ''
          Nucl.Phys. {\bf B518} (1998) 515-574}

\lref\gmp{B. R. Greene, D. R. Morrison, M. R. Plesser, ``Mirror
Manifolds in Higher Dimension,''
          Commun. Math. Phys. {\bf 173} (1995) 559-598}

\lref\mayr{P. Mayr, ``N=1 Superpotentials and Tensionless Strings on
Calabi-Yau Four-Folds,''
           Nucl.Phys. {\bf B494} (1997) 489-545}

\lref\lv{ N. C. Leung and C. Vafa, ``Branes and Toric Geometry,'' Adv.
Theor.Math.
Phys. {\bf 2} (1998) 91-118, hep-th/9711013}

\lref\ln{A. Lawrence and N. Nekrasov, ``Instanton sums and
five-dimensional theories,''
Nucl. Phys. {\bf B513} (1998) 93}

\lref\ckyz{T.-M. Chiang, A. Klemm, S.-T. Yau, and E. Zaslow,
``Nonlocal Mirror Symmetry:  Calculations and Interpretations,''
hep-th/9903053.}

\lref\bcovI{M. Bershadsky, S. Cecotti, H. Ooguri, and C. Vafa,
``Holomorphic Anomalies in Topological Field Theories,''
Nucl. Phys.  {\bf 405B} (1993) 279-304.}

\lref\bcovII{M. Bershadsky, S. Cecotti, H. Ooguri, and C. Vafa,
``Kodaira-Spencer Theory of Gravity and Exact
Results for Quantum String Amplitudes,''
Commun. Math. Phys. {\bf 165} (1994) 311-428.}

\lref\ionelpark{E.-N.- Ionel and T. Parker, ``Gromov-Witten
Invariants of Symplectic Sums,'' math.SG/9806013.}

\lref\wittwodgrav{E. Witten, ``Two-Dimensional Gravity and
Intersection Theory on Moduli Space,'' Surveys in Diff. Geom. {\bf 1}
(1991) 243-310.}

\lref\konttwodgrav{M. Kontsevich, ``Intersection Theory on the Moduli Space of
Curves and the Matrix Airy Function,''
Commun. Math. Phys. {\bf 147} (1992) 1-23.}

\lref\bateman{The
Bateman Project, {\sl Higher Transcendental Functions,} Vol. 1, Sec. 5.3.-5.6, 
A. Erdelyi ed., McGraw-Hill Book Company, New York (1953).}

\lref\kt{A. Klemm  and S. Theisen,
``Considerations of One-Modulus Calabi-Yau Compactifications: 
Picard-Fuchs Equations, K\"ahler
Potentials and Mirror Maps,'' Nucl. Phys. {\bf B389} (1993) 153-180.}  

\lref\agm{P. Aspinwall, B. Greene, and
D. R. Morrison, ``Measuring Small Distances in N=2 Sigma Models,'' 
Nucl. Phys. {\bf 420} (1994) 184-242.}

\lref\McKay{Brendan D. McKay, ``{\sl nauty} User's Guide''  available at 
http://cs.anu.edu.au/~bdm/ .}

\lref\wittenphasetransition{E.\ Witten, ``Phase Transitions in M-Theory and 
F-Theory,''
Nucl. Phys. {\bf 471} (1996) 195-216; hep-th/96031150 . }

%
\Title{\vbox{\hbox{hep-th/9906046}
\hbox{IASSNS-HEP--99/55}}}
{\vbox{\centerline{Local Mirror Symmetry at Higher Genus}}}
\vskip 0.2in
\centerline{Albrecht Klemm\footnote{$^{*}$}
{e-mail:  klemm@ias.edu} and Eric Zaslow\footnote{$^{**}$}
{e-mail:  zaslow@math.nwu.edu}}
\vskip 0.1in
\centerline{\it$^{*}$School of Natural Sciences, IAS,
Olden Lane, Princeton, NJ 08540, USA}
\centerline{\it$^{**}$Department of Mathematics, Northwestern
University,
Evanston, IL 60208, USA}

\vskip 0.3 in

\centerline{\bf Abstract}
\vskip 0.1in

We discuss local mirror symmetry for higher-genus curves.
Specifically, we consider the topological string partition
function of higher-genus curves contained in a Fano surface
within a Calabi-Yau.  Our main example is the local
${\bf P}^2$ case.  The Kodaira-Spencer theory of gravity,
tailored to this local geometry, can be solved to compute
this partition function.  Then, using the results of Gopakumar
and Vafa \vafagop\ and the local mirror map \ckyz,
the partition function can be rewritten in terms
of expansion coefficients, which are found to be integers.
We verify, through localization calculations in the A-model,
many of these Gromov-Witten predictions.  The integrality
is a mystery, mathematically speaking.  The asymptotic
growth (with degree) of the invariants is analyzed.
Some suggestions are made towards an
enumerative interpretation, following the
BPS-state description of Gopakumar and Vafa.

\Date{}

\newsec{Introduction and Summary}

In \ckyz, the techniques of mirror symmetry were applied
to the local geometry of a Fano surface inside a Calabi-Yau manifold.
Namely, from the data of the canonical bundle of the surface,
differential equations were proposed whose solutions
yield genus-zero, Gromov-Witten-type invariants.\foot{These
invariants are like the relative invariants of Ionel and Parker,
except that the degenerations are sort of ``one-sided.''}  From these, one
arrives
at integers after accounting for contributions from
multiple coverings.  The mirror symmetry predictions were
corroborated by independent A-model calculations using
localization.  The mirror principle was the main tool used
for verification; some more explicit fixed point calculations
were made, as well.
Excess intersection considerations
showed that the integers represent the
{\sl effective number} of rational curves ``coming from''
the Fano surface.

In this paper, we extend these techniques to higher-genus
curves.  For the A-model calculations, we use localization techniques on 
higher-genus moduli spaces to compute the Gromov-Witten invariants. The
mirror principle does not apply at higher genus. 
The analogue of the prepotential at higher genus
is the topological partition function at genus $g$, whose coefficients 
are the Gromov-Witten invariants.

For the B-model,
equations governing the dependence of these partition functions, 
$F^{(g)},$ on the complex structure moduli of the
Calabi-Yau were found in \bcovI \bcovII.  
In the effective 4-d N=2 supergravity theory, emerging from type IIB string theory 
compactification on the Calabi-Yau, the $F^{(g\ge 1)}$ appear as coefficients of 
the 
${\cal R}^2 {\cal F}^{2g - 2}$ terms (here ${\cal R}$ and ${\cal F}$
are the self-dual parts of the Riemann
tensor and the graviphoton field strength,
respectively).\foot{The genus zero prepotential $F^{(0)}$ 
determines the gauge kinetic terms and the masses of the 
charged BPS states. 
The information contained in the local geometry suffices to obtain 
the rigid theory (Seiberg-Witten theory) -- 
see \swrefs\ for reviews.}
The recursive nature of the equations comes from the fact
that Riemann surfaces with marked points degenerate
at the boundaries of moduli space -- either into pairs
of lower-genus
surfaces or surfaces with fewer marked points.
As the anti-holomorphic
dependence of the topological partition functions is
determined by an exact form, it only receives contributions from the boundaries.

One can tailor the methods of \bcovII\ to the local situation we
seek.  Our motivation for doing so depends on two
simplifications in the local case.  
First, the solution of the holomorphic anomaly equation using 
Kodaira-Spencer gravity simplifies considerably in the local 
case due to the fact that the propagators involving the descendents of 
the puncture operator (2-d dilaton) can be gauged to zero.  
Second, the higher-genus Gromov-Witten invariants are
computable using localization. This is in constrast to 
complete-intersection Calabi-Yau manifolds in toric varities, 
where the moduli space of higher-genus maps into the Calabi-Yau
is not simply expressible as the zero locus of a section of
a bundle over the moduli space of maps into an ambient toric
manifold.\foot{We thank R. Pandharipande for explaining this
to us.}  For Fano surfaces which are themselves toric, we need 
not cut with a section to obtain the moduli space of interest.
This implies in particular that
calculations can be made to fix the ambiguity in solving the 
Kodaira-Spencer recursions in the local case. 
In the global case of the quintic, by contrast, classical geometry cannot 
fix certain unknown constants.

After solving the Kodaira-Spencer theory and using the
mirror map, we verify mirror symmetry and apply the work of Gopakumar and Vafa 
\vafagop\
to organize the topological partition functions in terms
of coefficients which should be integers, as they count
BPS states of specified spin and charge (in the four-dimensional
theory, spin refers to a Lefschetz-type $SU(2)$
action on the BPS states).  This integrality check is
a strong test of the power of M-theory and the approach
of Gopakumar and Vafa.  The 
asymptotic growth of the invariants is also analyzed directly.

The mathematical interpretation of these integers is
still unclear.  A proper explanation seems to involve
moduli spaces of objects in derived categories.
Heuristic evidence exists in support of this observation.

Another explanation is lacking.  The form of the partition
functions predicted by Gopakumar and Vafa is a
generalization of the multiple cover formula $(1/k^3).$
The multiple cover formula is well-understood mathematically,
but the higher-genus generalizations are a mystery.

\vskip0.1in

We would like to note that Hosono, Saito, and Takahashi
\hst\ obtain BPS counts and compute
topological partition functions via B-model calculations
for the rational
elliptic surface, though their techniques are somewhat
different.  They achieve good corroboration of Vafa and
Gopakumar's formulas.  A-model calculations for this
surface would presumably be prohibitive.  Moore and Marino
also have achieved higher-genus results via heterotic
duality \marinomoore.

\newsec{Localization at Higher Genus}

The tools for the localization calculations have
been available since the work of Kontsevich \kontloc,
Li-Tian \litian,
Behrend-Fantechi \behfan, Graber-Pandharipande \locvir, and
Faber \faber.

The moduli space, $\overline{\cal M}_{g,0}(\beta;{\bf P}),$ of
stable maps from genus $g$ curves into some toric variety ${\bf P}$ with
image in
$\beta \in H_2({\bf P};{\bf Z})$ can have components of various
dimensions.  To do intersection theory on it, we need a fundamental
cycle of some ``appropriate'' dimension -- the virtual fundamental
class.

This procedure was developed in \behfan, and
the equivariant localization for spaces with perfect
deformation theories was developed in \locvir.
In that paper, the authors treat the case of interest to us:
the restrictions to the fixed point loci (under the
torus action) of the
equivariant fundamental cycle on the space of
holomorphic maps of genus $g.$  The authors
give explicit formulas for the weights.

In our examples, we are interested in calculating
Chern classes of $U_\beta,$ the bundle over moduli space
whose fiber at a point $(C,f) \in \overline{\cal M}_{g,0}(\beta;{\bf
P})$
is equal to $H^1(C,f^*K_{\bf P}).$  It is not difficult to
compute the equivariant Chern classes at the fixed loci.

Using these ingredients, we can perform localization calculations
at higher genus.   The Atiyah-Bott
fixed point formula -- true for virtual classes by \locvir\ -- reduces
the calculation to integrals over these fixed loci.
The different components of the fixed loci are represented
by different graphs.  We sum over the graphs using a computer
algorithm.
As all the higher-genus behavior must be centered at the fixed
points (there are no fixed higher-genus curves), the fixed loci
are all products of moduli spaces of curves, and the graphs
are decorated by genus data at each vertex as well as degree
data at each edge, as in \kontloc.  The integrals
over the moduli spaces are performed using the algorithm
of Faber \faber.

In the following subsections, we summarize this procedure.

\subsec{Localization Formulas}

As discussed in \ckyz, a Fano surface within a
complete-intersection Calabi-Yau
contributes to the Gromov-Witten invariants, producing an
effective ``number'' of rational curves.  In fact, the curves
are not isolated and there is
a whole moduli
space of maps into the Fano surface, arising as a
non-discrete zero locus of a section of a bundle
over the moduli space of maps in the Calabi-Yau.
Just as spaces of multiple coverings of isolated curves contribute
to the Gromov-Witten invariants by the excess intersection
formula,\foot{The contribution, from a component
of the zero locus $Y$ of some section,
to the Chern class $c(E)$ of some vector bundle $E$ over $M$ is
$\int_Y{c(E)\over c(N_{Y/M})},$ where $N_{Y/M}$ represents the
normal bundle of $Y$ in $M.$  The equation here follows from the
reasoning between eqs. (5.4) and (5.5) in \ckyz.}
the contribution from the surface $B$ can be calculated
as well.  It is
\eqn\locinv{K^g_\beta = \int_{\overline{\cal M}_{g,0}(\beta; B)}
c(U_\beta),}
where $\beta \in H_2(B;{\bf Z})$ and $U_\beta$ is the bundle whose
fiber over $(C,f) \in \overline{\cal M}_{g,0}(\beta; B)$ is the
vector space $H^1(C,f^*K_B).$\foot{Let $ev$ be the evaluation map
from $\overline{\cal M}_{g,1}(\beta; B)$ to $B$ and let $\pi$
be the map $\overline{\cal M}_{g,1}(\beta; B)\rightarrow
\overline{\cal M}_{g,0}(\beta; B)$ which forgets the marked point.
Then $U_\beta = {\cal R}^1\pi_* ev^* K_B.$}

We assume that $B$ has a toric description so that we may
use the torus action to define an action on ${\overline {\cal M}}_{0,0}(\beta; B)$ 
(moving the
image curves) and on $U_\beta,$ which inherits the natural
action on the canonical bundle.
We evaluate the Gromov-Witten invariants by localizing to
the fixed points using the Atiyah-Bott formula
$$\int_M \phi = \sum_{P} \int_P \left( i_P^*\phi \over e(\nu_P)
\right),$$
where the sum is over fixed point sets $P,$ $i_P$ is the embedding into
$M,$
and $e(N_{P/M})$ is the Euler class of the normal bundle of $P$ in $M.$

Graber and Pandharipande proved that this formula holds
mutatis mutandis, with the replacement of the integral
of a class by evaluation over virtual fundamental
cycles (in the equivariant Chow ring) of
equivariant classes of deformation complexes.
The equivariant Chern classes and bundles involved are then computed in
the standard way, e.g. by looking at the weights of
sections and taking alternating
products over terms in a complex.  In our case, following Kontsevich,
the
fixed moduli can be labeled by graphs.
The fixed locus ${\overline M}_{\Gamma}$ corresponding a graph $\Gamma$
is a product of
moduli spaces:
$${\overline M}_{\Gamma} = \prod_{v}{\overline{\cal M}_{g(v),val(v)}}.$$
Here $v$ are the vertices of the graph and $g(v)$ and $val(v)$
are the genus and valence of the vertex.  (At a vertex, which is
mapped to a torus fixed point, the choice of a $val(v)$-marked genus
$g(v)$
curve represents the only moduli.)

Graber and Pandharipande \locvir\ found
the weights of the inverse of the Euler class of the normal bundle to
be:
\eqn\virnorm{\eqalign{
{1\over e(N^{vir})}=
&\prod_{e}{(-1)^{d_e} d_e^{2 d_e}\over (d_e!)^2 
(\lambda_{i(e)}-\lambda_{j(e)})^{2 d_e}}
\prod_{{a+b=d_e\atop a,b\ge 0}\atop k\neq i(e),j(e)}
{{1\over {a\over d_e} \lambda_{i(e)}+{b\over d_e} \lambda_{j(e)}-\lambda_k}}\cr
&\times \prod_{v} \prod_{j\neq i(v)} \, 
(\lambda_{i(v)}-\lambda_j)^{val(v)-1}\cr 
&
\times \left\{\matrix{\displaystyle{\prod_{v}
\left[\left(\sum_{F} w_F^{-1}\right)^{val(v)-3}\prod_{F\ni v}
w_F^{-1}\right] }\hfil & {\rm if }\ g(v)=0 \cr 
\displaystyle{\prod_{v} \prod_{j\neq i(v)} \,
P_{g(v)}(\lambda_{i(v)}-\lambda_j,E^*) 
\prod_{F\ni v} {1\over w_F-e_F}   } & 
{\rm if} \ g(v)\ge 1 \, }
\right.  } }
where the case with $g(v)=0$ is from \kontsevich.
Here $i(v)$ is the fixed-point image of the vertex $v;$
edges $e$ represent invariant ${\bf P}^1$'s connecting the fixed
point labeled
$i(e)$ to $j(e);$ and flags $F$ are pairs $(v,e)$ of vertices
and edges attached to them.
$\omega_F = (\lambda_{i(F)}-\lambda_{j(F)})/d_e,$ where
$i(F) = i(e)$ and $j(F) = j(e)$ for the associated edge $e\in F.$
$e_F$ is the Chern class of the
line bundle over $M_\Gamma$ whose fiber is the point $e(F)\cap C_v,$
a ``gravitational descendant'' in physical
terms --
also known as a $\kappa$ class.
We have defined the
polynomial $P_g(\lambda,E^*) = \sum_{r=0}^{g}\lambda^r c_{g-r}(E^*),$
where $E$ is the Hodge bundle with fibers $H^0(K_C)$ on moduli space.

The bundle $U_\beta,$ whose top Chern class $\phi$ we are calculating, has
fibers $H^1(C,f^*K_B)$ over a point $(C,f).$  
Let us now fix $B = {\bf P}^2$ and put $\beta = d$ since
$H_2({\bf P}^2)$ is one-dimensional.
Since the
invariant maps are known and the maps from the genus $g(v) >0$
pieces are constants, we explicitly
calculate the weights of the numerator in the Atiyah-Bott formula
to be:
\eqn\numwts{i^*(\phi)=
\prod_{v}\Lambda_{i(v)}^{val(v)-1}P_{g(v)}(\Lambda_{i(v)},E^*)
\prod_{e}\left[\prod_{m=1}^{3d_e - 1}\Lambda_{i(e)} + {m\over d_e}
(\lambda_{i(e)} -\lambda_{j(e)})
\right],}
where $\Lambda_i = \lambda_1 + \lambda_2 + \lambda_3 - 3 \lambda_i.$

\subsec{Faber's algorithm}

We therefore have explicit formulas for the class to integrate
along $\overline{M}_\Gamma.$  The integrals involve the $\kappa$ classes and
the Mumford classes (from the Hodge bundle).
The integrals can be performed by the recursive algorithm developed
by Faber.

A rough sketch of the idea of Faber's algorithm is as follows
\faber.  Witten's
original recursion relations from topological gravity 
\wittwodgrav,
proved by Kontsevich \konttwodgrav,
suffice
to determine integrals of powers of the $\psi_i$'s, which are
the first Chern classes of the line bundles whose fiber is the
cotangent space of the corresponding marked point, $i = 1,...,n$
(the gravitational descendents).  However, the integrals over
$\overline{M}_{\Gamma}$ involve the $\lambda$ classes as well --
the Chern classes of the
Hodge bundle.  These classes can be represented by cycles
involving the boundary components
of moduli space as well as the $\psi_i$'s.

So a proper intersection theory including boundary
classes is needed.  The boundary components are images under maps from
$\overline{\cal M}_{g-1,n+2}$ (the map identifying the last two points;
this map is of degree two since swapping the points gives the same
boundary curve) and
$\overline{\cal M}_{h,s+1}\times\overline{\cal M}_{g-h,(n-s)+1}$
(the map identifying the two last points, which is degree one
except when both $n=0$ and $h=g/2$).  Understanding the pull-back
of the boundary classes under these maps completes the reduction
of intersection calculations to lower genera.  
For example, the $\psi_i$'s pull
back to $\psi_i$'s of corresponding points in the new moduli spaces.
One must also note that since the boundary divisors can intersect
(transversely, in fact), boundary classes must also be pulled back to
moduli spaces to which they do not correspond.  We refer the reader
to \faber\ for details.

What remains, then, is to express the
Chern classes of the Hodge bundle $E$
in terms of the boundary divisors.  This is
precisely the content of Mumford's
Grothendieck-Riemann-Roch (G-R-R) calculation.
Briefly (too briefly!), there is the
forgetful map $\rho: \overline{\cal M}_{g,1} \rightarrow \overline{\cal M}_{g,0}$
(we now set $\overline{\cal C}\equiv \overline{\cal M}_{g,1}$),
with respect to which we wish to push forward the relative
canonical sheaf $\omega_{\overline{\cal C},\overline{\cal M}}.$
Since $\rho_* \omega_{\overline{\cal C},\overline{\cal M}} = E,$
the G-R-R formula tells us the Chern character classes of $E$
(we need that $R^1\rho_* \omega_{\overline{\cal C},\overline{\cal M}}
= {\cal O}$).  The formula calls for integration of
$ch(\omega_{\overline{\cal C},\overline{\cal M}})$ with the Todd class
of the relative tangent sheaf.  The former involves the
descendents, while the relative tangent sheaf is dual
to the relative cotangent sheaf, which differs from
$\omega_{\overline{\cal C},\overline{\cal M}}$ only at the
singular locus (on smooth parts, sections of the canonical
bundle are just one-forms).  The Todd expansion is responsible for
the appearance of Bernoulli numbers, while the 
difference from the canonical sheaf introduces the divisor
classes.

Therefore, integrals of descendent classes and $\lambda$ classes
can be performed by using Mumford's formula
(see section 1 of \fabpan)
to convert $\lambda$ classes
to descendents and divisor classes, then on the boundaries of
$\overline{\cal M}$ pulling these clases back to the moduli spaces which
cover the boundaries.  As these spaces have lower genera, the
integrals can be performed recursively.  Faber has written a computer 
algorithm in Maple for this procedure, which he generously lent 
us for this calculation.

\subsec{Summing over graphs}

The contribution to the free energy $F^{(g)}$ from genus $g$ curves in the 
class $\beta$ is given from the fixed point formula as
\eqn\application
{K_\beta^g=\sum_{\Gamma} {1\over |{\bf A}_\Gamma|}\int_{{\overline M}_\Gamma} 
{i^*(\phi)\over e(N^{vir})}\ ,}
with $e(N^{vir})^{-1}$ and $i^*(\phi)$ as described in \virnorm\numwts. 
The formal expansion of the integrand 
yields cohomology elements of ${\overline {\cal M}}_{g,n}$, the moduli space of 
pointed 
curves at the vertices. The integral over ${{\overline M}_\Gamma}$ splits into 
integrals 
over these moduli spaces, against which the cohomology elements of the right 
degree have to be 
integrated. In addition, the contribution of a given graph must be divided by the 
order of 
the automorphism group ${\bf A}_\Gamma$, as we are performing intersection theory 
in the orbifold sense. ${\bf A}_\Gamma$ contains the 
automorphism group of $\Gamma$ as a marked graph and a ${\bf Z}_{d_e}$ factor for 
each edge which maps with degree $d_e$.    
  
Hence it remains to construct the graphs $\Gamma$ which label the fixed
point loci of 
${\overline {\cal M}}_{g,k}(d,\IP^2)$ under the induced torus action,
then to carry out the summation and 
integration.  A graph is labeled by a set of vertices and edges, and can
be viewed as a degenerate domain curve with additional data specifying the 
map to $\IP^2$. The vertices $v$ represent the irreducible components 
$C_{v}$. They are mapped to fixed points in $\IP^2$ under the torus action. 
The index of this fixed point $i(v)$ and the arithmetic genus $g(v)$ of 
the component $C_v$ are additional data of the graph. 
The edges $e$ represent projective lines which are mapped invariantly, with degree 
$d_e,$ to projective lines in the toric space connecting the pair of 
fixed points. A graph without the additional data $g(v),d_e,i(v)$ will be 
referred to as {\sl undecorated}.

The following combinatorial conditions 
specify a $\Gamma_{g,d,k}$ graph representing a
fixed-point locus of genus 
$g,$ degree $d$ maps with $k$ marked points:\foot{Condition
four is relevant only for maps with $k$ marked points, which we do
not consider $(k=0).$ $S(v)$ is then the subset of marked points on the 
component $C_v$.}

\noindent 
(1) if $e\in {\rm Edge}(\Gamma)$ connects the vertices $(u,v)$ then
$i(u)\neq i(v)$

\noindent
(2) $1-\#({\rm Vertices} )+\#({\rm Edges})+\sum_{v\in {\rm Vert}(\Gamma)} g(v)=g$      

\noindent
(3) $\sum_{e \in {\rm Edge}(\Gamma)} d_e=d$ 

\noindent 
(4) $ \{1,\ldots , k\}=\cup_{v\in {\rm Vert}(\Gamma)} S(v).$

To construct all $\Gamma_{g,d,0}$ graphs we start by generating a list of all 
possible 
undecorated graphs that can be decorated to yield $\Gamma_{g,d,0}$ graphs. 
Hence the number of edges is restricted by (3) while the number 
of vertices is restricted by (2). The undecorated graphs with $n_v$ vertices can 
be 
represented by an $n_v\times n_v$
symmetric incidence matrix, $m$. Entries $m_{i,j}=1$ or $0$ 
represent a link or no link between $v_i$ and $v_j,$
so with a fixed number of edges $n_e$
there will be $\left({{(n_v+1)n_v\over 2}\atop n_e}\right)$ 
possibilities.  A main problem is to identify among the graphs
those which have different 
topology. To obtain invariant data, independent of the ordering of the 
vertices and depending only on the topology of the graph, 
we use a so called depth-first-search algorithm.  This starts with a
vertex $v_k$ and descends level-by-level along all occuring branches 
collecting the data of the encountered vertices, namely the valence 
(and in later applications the additional information $g(v)$ and $i(v)$
as well as edge data 
$d_\alpha$ for the map), into a list $T_k$ graded by the level, until all 
vertices have been encountered.  A lexicographic ordering 
can be imposed on the combined lists of all vertices 
$T_\Gamma=\{T_1,\ldots,T_{n_v}\}$ 
and  
$T_{\Gamma_i}=T_{\Gamma_k}$ only if $\Gamma_i$ is isomorphic to 
$\Gamma_{k}$. These
invariants will be used for a.) generating distinct undecorated
graphs, b.) decorating them without redundancy, and c.) finding the 
automorphism group.\foot{A useful computer program \McKay\  was used 
to check the number of undecorated graphs.}  Fig. 1 shows as an example the 
generation 
of genus 2, degree 3 graphs.

\vskip 0.5cm
{\baselineskip=12pt \sl
\goodbreak\midinsert
\centerline{{ \epsfxsize 5.2 truein\epsfbox{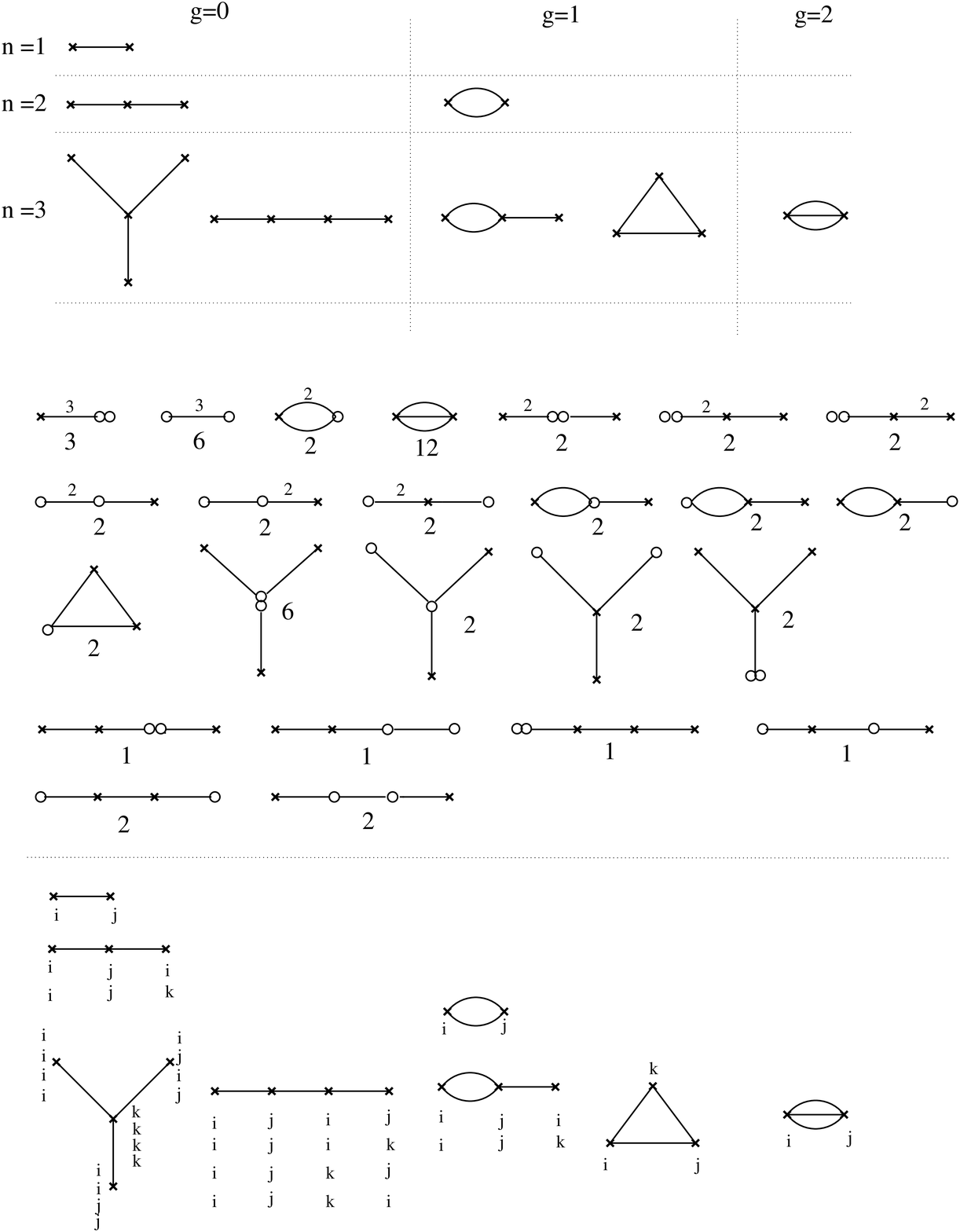}}}
\leftskip 1pc\rightskip 1pc \vskip0.3cm
\noindent{\ninepoint  \baselineskip=8pt  
{\bf Fig. 1:} The first part of the figure shows a scheme of all undecorated 
graphs
that can be decorated  
to $g=2,$ $d=3$ graphs. In the middle we show the $24$ different ways to decorate 
them by 
$d_e$ (small numbers, if different from one) and $g(v)$ (number of circles), with 
the order of the 
automorphism ${\bf A}_\Gamma$ indicated by the bigger numbers. 
In the last part of the figure we show the decoration possibilities by the 
$i(v)$ indices, which lead to $63$ graphs. The indices $(i,j,k)$ are finally 
summed 
over all permutations of $(1,2,3)$. }
\endinsert}

We have implemented the above generation of graphs\foot{The number of graphs grows
very quickly with the genus.  E.g., to calculate the $d=5$ term 
(for $g=0,\ldots,5$) in (2.5), one sums over six times 
$173\{21\},$ $733\{101\},$ $2295\{313\},$ $5353\{719\},$ $11101\{1442\},$ 
$20345\{2570\}$  
fully-decorated graphs, 
where the number in braces indicates the number of graphs with just 
$g(v)$ and $d_e$ decorations.  For example, in the calculation of $(g,d)=(4,5)$ 
the 
maximal dimension of a vertex moduli space ${\rm dim}(\overline{{\cal 
M}}_{g,n})=14$
occurs in the ``star graphs'' -- irreducible genus 4 curves with five legs. 
At this dimension, $22462$ different 2-d gravity integrands must be evaluated.}
of \application~in a completely automated 
computer algorithm that uses Faber's algorithm to perform the integrals over 
${\overline {\cal M}}_{g,n}$.  We exhibit the results 
of these localization calculations in
the instanton pieces $F^{(g)}_{inst} = \sum_{d>0}
K^g_d q^d$ of the partition functions $F^{(g)}$ listed below: 

\eqn\fspii{\eqalign{
F^{(0)}&=-{\frac{{t}^3}{18}}+ 3\,q - {\frac{45\,{q^2}}{8}} + 
{\frac{244\,{q^3}}{9}} - 
  {\frac{12333\,{q^4}}{64}} + {\frac{211878\,{q^5}}{125}}\ldots \cr
F^{(1)}&={-\frac{t}{12}} + {\frac{q}{4}} - 
  {\frac{3\,{{q}^2}}{8}} - {\frac{23\,{{q}^3}}{3}} + 
  {\frac{3437\,{{q}^4}}{16}} - 
  {\frac{43107\,{{q}^5}}{10}}\ldots \cr
F^{(2)}&=\frac{\chi}{5720}+ {\frac{q}{80}} + 
  {\frac{3\,{{q}^3}}{20}} - 
  {\frac{514\,{{q}^4}}{5}} + 
  {\frac{43497\,{{q}^5}}{8}}\ldots\cr
F^{(3)}&=-
\frac{\chi}{145120}+\frac{q}{2016} + {\frac{{q^2}}{336}} + 
  {\frac{{q^3}}{56}} + 
  {\frac{1480\,{q^4}}{63}} - 
  {\frac{1385717 \,{q^5}}{336}}\ldots\cr
F^{(4)}&=\frac{\chi}{87091200}+\frac{q}{57600} + 
  {\frac{{q^2}}{1920}} + 
  {\frac{7\,{q^3}}{1600}}  -{\frac{2491\, q^4}{900}} + 
  {3865234\,{q^5}\over 1920} \ldots \cr 
F^{(5)}&=-\frac{\chi}{2554675200} +\frac{q}{1774080}+\frac{q^2}{14080}+\frac{61\, 
q^3}{49280}+ 
\frac{4471\, q^4 }{22176}-
\frac{65308319\, q^5}{98560}\ldots }}

\subsec{Organizing the partition functions}

In order to make enumerative predictions from the partition
functions, we need
the analogue of the multiple-cover formula ($1/k^3$). 
In other words, we need to know
how a ``fundamental object,'' a holomorphic
curve or D-brane (BPS state) of given
charge and spin content, contributes to the partition function.

The functional form of $F^{(g)}$ was derived in \vafagop:
\eqn\gcovers{F(\lambda) = \sum_{g=0}^{\infty}F^{(g)}\lambda^{2g-2}
= \sum_{\{d_i\}, g\geq 0,  k>0}
n^g_{\{d_i\}}{1\over k}(2 \sin {k\lambda\over 2})^{2g-2}
{\rm exp}[-2\pi k \sum_i d_i t_i].}
Here the $\{ d_i\}$ define the homology class (charge) of
the BPS state, and $n^g_{\{d_i\}}$ are the number of
BPS states of charge $\{ d_i\}$ and left-handed spin
content described by $g.$\foot{BPS states are killed by
the right-handed supersymmetry generators (half), where
$so(4) \cong su(2)_L \otimes su(2)_R.$  BPS supermultiplets,
formed from a spin $(j_1,j_2)$ Fock ``vacuum,'' have states
with left-handed content $[(\half)\oplus 2(0)]\otimes j_1,$
while the $2j_2 + 1$ right-handed states only contribute an overall
factor (with sign).  Defining $I = (\half)\oplus 2(0),$
$g$ labels the representation $I^{\otimes g}.$ }

The table below was generated by extracting the integers from
the partition functions as indicated.  Results for degrees
higher than five come from the B-model calculation, which
we review in the next section.

\vskip 0.2in

{\vbox{\ninepoint{
$$
\vbox{\offinterlineskip\tabskip=0pt
\halign{\strut
\vrule#&
&\hfil ~$#$
&\hfil ~$#$
&\hfil ~$#$
&\hfil ~$#$ 
&\hfil ~$#$
&\hfil ~$#$
&\vrule#\cr
\noalign{\hrule}
&d
&g=0
&g=1
&g=2
&g=3
&g=4
&
\cr
\noalign{\hrule}
&1
&3
&0
&0
&0
&0
&\cr
&2
&-6
&0
&0
&0
&0
&\cr
&3
&27
&-10
&0
&0
&0
&\cr
&4
&-192
&231
&-102
&15
&0
&\cr
&5
&1695
&-4452
&5430
&-3672
&1386
&\cr
&6
&-17064
&80948
&-194022
&290853
&-290400
&\cr
&7
&188454
&-1438086
&5784837
&-1536990
&29056614
&\cr
&8
& -2228160
&25301295
&-155322234 
& 649358826
&-2003386626
&\cr
&9
& 27748899
&-443384578
&3894455457
&-23769907110
&109496290149
&\cr
&10
&-360012150
&7760515332
&-93050366010
&786400843911
&-5094944994204
&\cr
\noalign{\hrule}}\hrule}$$}
\centerline{{\bf Fig. 2:} The (weighted)
number of BPS states $n^g_d$ for the local ${\bf P}^2$ case.}
\vskip7pt}

As an immediate check, we note that $n^g_d = 
(-1)^{{\rm dim}({\cal N}_d)} (d+1)(d+2)/2$ when $g = \pmatrix{d-1\cr 2}.$
In the BPS language, these integers count highest-spin two-brane states
with a given charge.\foot{This case was first
discussed in \wittenphasetransition.}  The moduli space of polynomials of
degree $d$ is ${\cal N}_d\equiv {\bf P}^{(d+1)(d+2)/2 - 1},$ and
$(d+1)(d+2)/2$ equals the Euler characteristic of this
space.  A simplification for highest-spin
states allows us to ignore the moduli corresponding to
bundles on D-branes \vafagop. 

If for a given degree $d$ all non-vanishing $n_d^g$ are known,
one can reorganize the spin 
content of BPS states with a fixed charge in the irreducible
representations of the left 
$su(2)_L$. With $n^5_5=-270$ from \fspii\ and $n^6_5=21$ for the smooth genus six
curve, we get
$$
\vbox{\offinterlineskip\tabskip=0pt
\halign{\strut
\vrule#&
&\hfil ~$#$
&\hfil ~$#$
&\hfil ~$#$
&\hfil ~$#$ 
&\hfil ~$#$
&\hfil ~$#$
&\hfil ~$#$ 
&\hfil ~$#$
&\hfil ~$#$
&\vrule#\cr
\noalign{\hrule}
&d
&s=0
&s={1\over 2}
&s=1
&s={3\over 2}
&s=2
&s={5\over 2}
&s=3
&s={7\over 2}
&
\cr
\noalign{\hrule}
&3
&4
&-13
&-10
&0
&0
&0
&0
&0
&\cr
&4
&-27
&24
&24
&18
&15
&0
&0
&0
&\cr
&5
&78
&-57
&12
&-12
&-12
&36
&24
&21
&\cr
\noalign{\hrule}}\hrule}$$
Note that in this basis, the numbers are considerably
smaller.  It is curious that the next-to-last number in each
row is three greater (in absolute value) than the last.
Further enumerative
remarks are made in section four. 

The form \gcovers\ amounts to a set of multiple-cover formulas
which can be found by extracting the appropriate coefficient
of $\lambda$ and $k.$  Let us denote by $C_g(h,d)$ the contribution
of a genus $g$ object in a class
$\beta$ to the partition function at genus $g+h$ in class $d\beta.$
>From \gcovers\ we see that $C_g(h,d)$ is the coefficient
of $\lambda^{2(g+h)-2}$ in $d^{-1}(2\sin (d\lambda/2))^{2g-2}.$
Some of these numbers have been corroborated
by mathematical calculations.
For example, $C_0(0,d) = 1/d^3$
\aspmor\manin .  $C_0(h,d)$ was calculated in
\fabpan, and $C_g(h,1), g\geq 1$ in \pandh.
R. Pandharipande has communicated to us another subtraction
scheme when $h=0$ which also yields integers.

A puzzle remains, however.
The multiple-cover formulas predict, for example,
that genus two objects of degree four
contribute to the partition function at genus two, degree
eight.  However, no double covers of genus two curves by
genus two curves are expected.
In general, since no covering maps from genus $g$
to $g+h$ curves exist, how does one even {\sl define} the covering
contribution?  Perhaps a formulation via spaces of sheaves (which we
discuss in section four) is in order.

\subsec{Example}

For example, the coefficient $K^2_4 = -514/5$ of $q^4$ in $F^{(2)}$
in \fspii\ can receive contributions from the non-zero
integer invariants $n^0_1,$ $n^0_2,$ $n^0_4,$
$n^1_4,$ and $n^2_4.$  Since $C_0(2,d) = d/240,$
$C_1(1,d) = 0,$ and $C_2,(0,d) = d,$
we see
$$K^2_4 = 3 (4/240) - 6 (2/240) - 192 (1/240) + 231 (0) - 102 (1)
= -514/5.$$

\newsec{B-Model Calculations}

Though \bcovII\ developed recursion relations for Calabi-Yau
threefolds, the lesson of \ckyz\ is that considerations
involving holomorphic curves inside Fano subvarieties within
Calabi-Yau threefolds may be isolated to calculations on the
canonical bundle qua normal bundle of the surface.  The reason
for
this is that all curves within the surface have negative
self-intersection and so can't be moved off of the surface.
The topological partition functions count holomorphic
curves of higher genus, in some sense, so we expect to
find a ``local'' version of the
equations of Kodaira-Spencer gravity,
developed in \bcovI\bcovII. We will now describe the 
simplifications in the Kodaira-Spencer theory
as we specialize to the local case.

\subsec{The vacuum line bundle and correlation functions}

In $N=2$ topological theories the vacuum state $|0\rangle$ transforms
as a section of a holomorphic line bundle $\cal L$ over the moduli 
space ${\cal M}$ of the theory. In the case of the B-type topological 
$\sigma$-models, the former will be identified with the holomorphic 
(3,0)-form $\Omega$ while ${\cal M}$ will be identified with the complex 
structure moduli space of the Calabi-Yau geometry $\hat X$, which is mirror dual 
to 
the geometry $X$ on which we count the holomorphic maps. $\cal M$ has a special
K\"ahler structure, with K\"ahler potential $K(z,\bar z)$. 

By the sewing axioms of topological field theory the vacuum amplitudes 
at genus g $F^{(g)}$ are sections of ${\cal L}^{2-2g}$ over $\cal M$. 
The  general topological correlations function $F^{(g)}_{i_1\ldots i_n}\in 
{\cal L}^{2-2 g} \otimes {\rm Sym}_n T^* {\cal M}$  are defined as 
covariant derivatives of the ``potentials''
$F^{(g)}$ as $F^{(g)}_{i_1\ldots i_n}=D_{i_1}\ldots D_{i_n} F^{(g)}$.
The $D_{i}$ are covariant with respect to
metric connection $\Gamma_{lk}^i=G^{i\bar m}\partial_l G_{k\bar m}$ 
of the Weil-Peterson-Zamolodchikov metric $G_{k\bar l}(z,\bar z)=\partial_k{\bar 
\partial}_{\bar l} K$ 
on ${\cal M}$ as well as the K\"ahler connection $A_i=\partial_i K$ of the line  
bundle ${\cal L}$. 

\subsec{The key holomorphic object at genus  zero:  $F^{(0)}_{i_1,i_2,i_3}$} 
  
The genus zero three-point functions $F^{(0)}_{i_1,i_2,i_3}$, physically 
well-known as the Yukawa couplings of the heterotic string with standard
embedding or the 
magnetic moments in the type-II compactifications, are of special importance 
because, as their moduli space has no boundaries, they are truly holomorphic 
and appear as building blocks in the recursive definition of the higher loop
amplitudes. 
 
They can be explicitly calculated as integrals over the Calabi-Yau manifold  
$\hat X$ 
\eqn\yuk{F^{(0)}_{i_1,i_2,i_3}(z)=-\int_{\hat X} \Omega \, 
\partial_{z_{i_1}} \partial_{z_{i_2}}\partial_{z_{i_3}}\Omega }  
which can be expressed as the derivatives of suitable combinations
of period integrals over the the cycles of $\hat X$.  These period integrals 
can be most easily reconstructed from solutions of the Picard-Fuchs system of 
$\hat X$. For the local case, the $F^{(0)}_{i_1,i_2,i_3}$ can
be defined either as a limit of a global Calabi-Yau geometry or intrinsicially 
from the Picard-Fuchs system for the local geometry.  We sometimes
write $C_{i,j,k}$ for $F^{(0)}_{i,j,k}.$

\subsec{The canonical coordinates $t_i$ and the holomorphic limit}

Other correlation functions $F^{(g)}_{i_1,\ldots, i_r}$ for $g>0$ are
not holomorphic.\foot{Note that 
$F^{(0)}_{i_1,\ldots,i_n}$ for $n=0,1,2,$ may have an anti-holomorphic 
dependence.} 
However, due to a uniquely distinguished class of coordinate systems on ${\cal 
M}$, 
the {\sl canonical}
coordinates, there is a well-defined  limit one can take to define 
holomorphic correlators. 

Of the $2h^{2,1}+2=2{\rm dim}({\cal M})+2$ period integrals
$\omega_i=\int_{\gamma_i} \Omega$, 
near the maximal unipotent point
$P_M = (z_i=0, {\rm Im}(t_i)\rightarrow \infty ) \in \cal M,$
exactly $h^{2,1}$ will have logarithmic behavior
$\omega_i\sim \log(z_i)+O(z),$ $i=1,\ldots,h^{2,1}.$ 
A unique one $\omega_0=1+O(z)$ is analytic. 
The homogeneous coordinates, defined as $2 \pi i \ t_i=
{\omega_i\over \omega_0} \sim \log(z_i),$ have the following 
properties, which define canonical coordinates near any point $P_0$.

All holomorphic derivatives vanish at $P_0$
\eqn\canonical{\partial_{t_1}\ldots \partial_{t_r}\Gamma_{ij}^k|_{P_0}=0, \quad
\partial_{t_1}\ldots \partial_{t_r}K|_{P_0}=0\ .} 
Let as above $P_0$ be at $z=0$, $t=t_0$. Then \canonical\ 
implies that in the $t$ coordinates the leading term in $\bar \lambda_i =(\bar 
t_i-\bar t_{0})$ of  
$K=C + O({\bar \lambda })$ and $G_{i\bar j}=C_{i,\bar j}+O(\bar \lambda)$ is 
constant. 
When re-expressed in the coordinates $z_i$, the holomorphic parts of $K$ and $G$ 
in 
the $\bar \lambda_i\rightarrow 0$
limit are
\eqn\limit{K=C-\log(\omega_0),\qquad G_{i\bar m}={\partial t_k\over \partial z_i} 
C_{k\bar m}\ . }
In all quantities to be discussed below we will 
take the holomorphic $\bar \lambda\rightarrow 0$ limit.

The $t_i$ at $P_M$ have the additional
property that they are identified by mirror symmetry
with the complexified K\"ahler parameters of the mirror $ X,$
and the $F^{(g)}_{i_1,\ldots, i_r}$ 
are curve-counting functions for holomorphic maps of genus $g$ into $X$. 
Note also that $t_i$ at $P_M$ are 
the special coordinates in which the holomorphic
parts of the K\"ahler connection and the 
Weil-Peterson connection vanish, i.e. $D_i$ becomes
the ordinary derivative $\partial_{t_i}$.

At this point there is important simplification in the local case. 
As is clear
from the differential equations associated to the
local case \kkv \ckyz, $\omega_0=1$ is 
always a holomorphic solution, and hence for the local case the holomorphic part 
of the 
K\"ahler potential becomes trivial in the limit \limit .   

\subsec{The holomorphic anomaly at genus one and the derivation of $F^{(1)}$}

The genus one topological partition function is defined as 
\eqn\fgone{F^{(1)}={1\over 2}\int_{\cal F} {\dd^2 \tau\over
{\rm Im}(\tau)} {\rm Tr}_{{\rm Ra},
{\overline {\rm Ra}}}\, (-1)^F 
F_L F_R q^{H_L} {\bar q}^{H_R}\ . }
Note that without the insertions of the left and right fermion number operators, 
the integrand would be the Witten index and just receive contributions 
from the ground states ${\rm Tr}(-1)^F=\chi$. The holomorphic
anomaly due to contributions
from the boundary of the moduli space was derived in \bcovI\ for
general $N=2$ theories 
and specializes for the $\hat c=3$ case to\foot{The normalization
of $F^{(1)}$ in \bcovI\ is by a 
factor $2$ greater than the one in \bcovII . We will adopt the latter. }
\eqn\holanomaly{\partial_i\bar \partial_{\bar j}F^{(1)}={1\over 2} F^{(0)}_{ijk} 
{\bar F}^{(0)}_{\bar j\bar k\bar l} 
e^{2 K}G^{j \bar k} G^{k\bar l}-\left({\chi\over 24}-1\right)G_{i\bar j}\ .}
Using $R_{i\bar j}= -{1\over 2} {\partial^2\log
\det(G) \over \partial_i \bar \partial_{\bar j}}$ and the
special geometry relation 
\eqn\specialgeom{R_{i\bar j l}^k=G_{i\bar j}\delta^k_l+G_{l\bar j} \delta^k_i-
F^{(0)}_{ilm}{\bar F}^{(0)}_{\bar j\bar p\bar q} e^{2 K} G^{k\bar p} G^{m\bar q}\ 
,}
one can integrate \holanomaly\ up to an unknown holomorphic function $f$ to yield
\eqn\fI{F^{(1)}=\log\left(\det(G^{-1})^{1\over 2}
e^{{K\over 2}(3+h^{2,1}-{1\over 12} \chi)} |f^2|\right)\ . }

The holomorphic ambiguity $f$ can be parameterized by the
vanishing or pole behavior at the
discriminant loci $f=\prod_{i=1}^k (\Delta_k)^{r_i} \prod_{i=1}^{h^{2,1}} 
z_i^{x_i}$. 
In particular, the 
$x_i$ can all be determined by the limiting
behavior $\lim_{z_i\rightarrow 0} F^{(1)}=-{1\over 24} 
\sum_{i=1}^{h^{2,1}} t_i \int_X c_2 J_i$. The
behavior of $F^{(1)}$ at certain types of singularities
is universal -- e.g., for the conifold singularity
$r_{con}=-{1\over 12}$. By virtue of the remark at 
the end of the previous section, $\chi$ and $h^{1,2}$ only affect
an irrelevant additive constant to $F^{(1)}$ 
in the local case. Note that on the other hand, the
topological number $\int_{X} c_2 J_i$ is crucial 
for fixing the large $t_i$ behaviour of $F^{(1)}$ in the local case. 
Formulas for $\int_X c_2 J_i$ in the local case can be found in \ckyz .

\subsec{The higher genus anomaly and the derivation of the  $F^{(g)}$}
  
The higher genus correlators are defined by a recursive procedure using the 
holomorphic 
anomaly equation
\eqn\rec{\eqalign{\bar \partial_{\bar i} F^{(g)}_{i_1,\ldots,i_n}&={1\over 2} 
{\bar C}^{jk}_{\bar i} F^{(g-1)}_{jk i_1,\ldots, i_n}+\cr &{1\over 2}
{\bar C}^{jk}_{\bar i}\sum_{r=0}\sum_{s=0}{1\over s!(n-s)!}\sum_{\sigma\in S_n} 
F^{(r)}_{ji_{\sigma(1)}\ldots i_{\sigma(s)}}  
F^{(g-r)}_{ki_{\sigma(s+1)}\ldots i_{\sigma(n)}}\cr
&-(2 g -2 + n-1)\sum_{s=1}^n G_{\bar i,i_s} F^{(g)}_{i_1 \ldots i_{s-1} 
i_{s+1}\ldots i_n \ ,}}}
where ${\bar C}^{ij}_{\bar i}\equiv \bar F^{(0)}_{\bar i\bar j\bar k}
e^{2 K} G^{j\bar j}G^{k\bar k}$.
The right-hand side corresponds to the boundary contributions of the moduli space  
of marked Riemann surfaces, $\overline{{\cal M}}_{g,n}.$ The first term
comes from pinching a handle, the next from 
splitting the surface into two components by growing a long tube,
the last from when the insertion approaches marked points. 

The solution of the equation is provided by the calculation of potentials
for the non-holomorphic quantities ${\bar C}^{ij}_{\bar i}$. 
In the first, step one calculates $S^{ij}\in {\cal L}^{-2} \otimes 
{\rm Sym}_2 T {\cal M}$  such that
${\bar C}^{ij}_{\bar i}={\bar \partial}_{\bar i}S^{ij}$. 
This follows from \specialgeom\ by noting that in K\"ahler geometry $R^k_{i{\bar 
j} l}
=-{\bar \partial}_{\bar j} \Gamma^k_{il},$ and hence
\eqn\sij{{\bar \partial}_{\bar k}[S^{ij}C_{jkl}]=
{\bar \partial}_{\bar k}[\delta^i_l \partial_k K+
\delta^i_k\partial_l K+\Gamma^i_{kl}]\ . }
The derivatives $\bar \partial_{\bar k}$ on both sides can be
removed at the price of introducing a
meromorphic object $f^i_{kl}$, which also has to
compensate the non-covariant transformation properties
of quantities on the right-hand side. 
{\baselineskip=12pt \sl
\goodbreak\midinsert
\centerline{\epsfxsize 5 truein\epsfbox{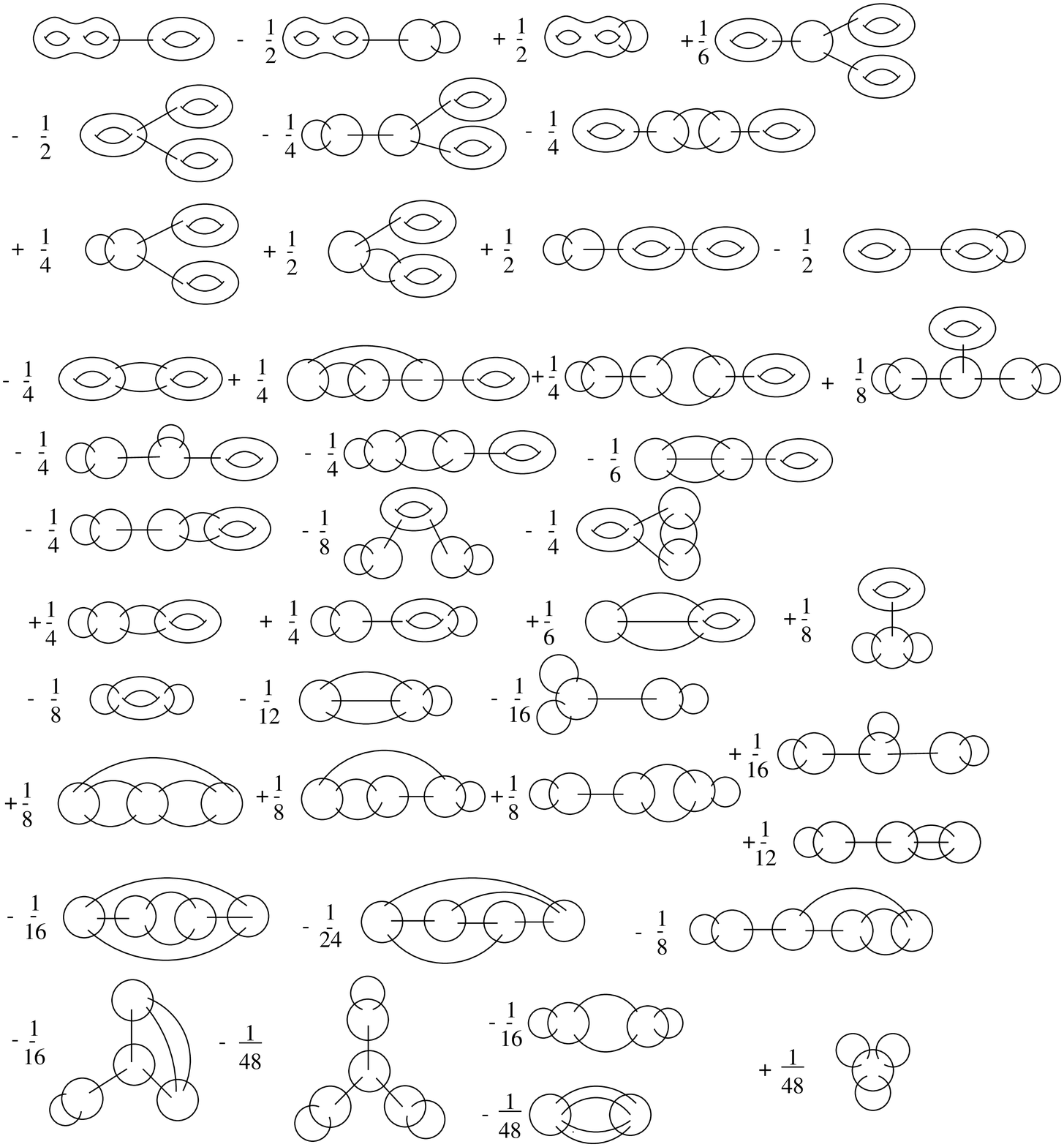}}
\leftskip 1pc\rightskip 1pc \vskip0.3cm
\noindent{\ninepoint  \baselineskip=8pt  
{\bf Fig. 3:} The degenerations of a genus 3 Riemann surface, with the symmetry
factors by which they contribute to the calculation of $F^{(3)}$.  In the local 
case, these are
all the contributions.  The diagrams correspond (in order) to terms in the
expression for $F^{(3)},$ though one expression may involve several
graphs.  For example, the fifth line in the figure contains the three
graphs which contribute to $-{2\over 3} S_2^4 
F^{(1)}_{,1}F^{(0)}_{,4}F^{(0)}_{,3}.$
Lines denote factors of $S_2,$ genus $g$ components with valence $n$
represent factors of $F^{(g)}_{,n}.$
In the global case, there are 81 additional graphs with $\phi$ propagators. 
}
\endinsert}

Therefore it is natural to split $f^i_{kl}$ into quantities with
simple transformation properties:
$f^i_{kl}=\delta^i_k \partial_l \log f+\delta^i_l \partial_k \log f-
v_{l,a} \partial_k v^{i,a}+\tilde f^i_{kl}$, where $\tilde f^i_{kl}$ now 
transforms
covariantly, $f\in {\cal L},$ and $v^{i,a}$ transform as tangent vectors. 
The choices of the 
$f$, $v^{i,a}$, $\tilde f^i_{kl}$ are by no means independent, however. 
If we specialize to the one-modulus case, as in our main example the local 
$\IP^2$,
we can in addition set
$\tilde f^z_{zz}=0$ and in the holomorphic limit get  the simplified expression
\eqn\sprop{\eqalign{S^{zz}&={1\over F^{(0)}_{zzz}}
\left[2 \partial_z \log(e^K |f|^2)-(G_{z\bar z}v)^{-1} \partial_z(v G_{z\bar 
z})\right]\cr 
&= -{1\over F^{(0)}_{zzz}}\partial_z \log\left(v {{\partial t}\over {\partial 
z}}\right)
\qquad {\rm as }\; \ \overline{\lambda}\rightarrow 0, }}
where $v\in T{\cal M}$ to render $S^{zz}$ covariant. The simplification in the 
second line is due to the fact that $K$ is constant in the holomorphic limit 
for the local case and $f$ can chosen to be constant as well.
Further potentials needed to solve for the $F^{(g)}$ in the global cases are 
$S\in {\cal L}^{-2},$ with $C_{\bar j \bar k \bar l}=e^{-2 K} D_{\bar i} D_{\bar 
j} 
\bar \partial_{\bar k} \bar S$, $S_{\bar i}\equiv \bar \partial _{\bar i} S$,
$\bar \partial_{\bar i} S^j=G_{i\bar i}S^{ij}$.  $K^{ij}=-S^{ij}$, 
$K^{i\phi}=-S^{i}$ and 
$K^{\phi\phi}=-2 S$ can be interpreted as propagators in the topological gravity 
theory, where $\phi$ is the dilaton, the first descendent of the puncture 
operator.   
However, in the local case we will see that a choice 
of the different holomorphic ambiguities can be made so that 
$K^{i\phi}$ and $K^{\phi\phi}$ vanish. $S^i$ is derived from
$${\bar \partial}_{\bar z}S^z={1\over F^{(0)}_{zzz}}{\bar \partial}_{\bar z}
\left[2 \partial_z \log(e^K |f|^2)^2-v^{-1} \partial_z(v \partial_z K)\right],$$
and if we set the holomorphic ambiguity in the special solution to this equation 
to zero, it vanishes in the holomorphic limit for the local case. 
Similarly one can see that $S$ vanishes.

The derivation of $F^{(g)}$ proceeds recursively. 
One first considers the holomorphic anomaly equation of $F^{(g)},$ 
and using ${\bar C}^{ij}_{\bar i}={\bar \partial}_{\bar i}S^{ij},$
one can write the right-hand 
side e.g. for $g=2$ as ${1\over 2}{\bar \partial}_{\bar i} 
[S^{jk}(F^{(1)}_{jk}+F^{(1)}_j F^{(1)}_k)]-
{1\over 2}S^{jk}{\bar \partial}_{\bar i} [F^{(1)}_{jk}+F^{(1)}_j F^{(1)}_k]$. 
Using the definition of the
Riemann tensor as commutator and special geometry, i.e. 
$[{\bar \partial}_{\bar i},D_j]_k^l=
-G_{\bar i j}\delta^l_k-G_{\bar i k}\delta^l_j+C_{jkm}\bar C^{ml}_{\bar i},$ 
one lets the ${\bar \partial}_{\bar i}$ derivative act on $F^{(g-1)}$ and repeats 
the procedure until
an expression ${\bar \partial}_{\bar i}F^{(g)}={\bar \partial}_{\bar i}[\ldots]$ 
is derived, where $[\ldots]$
contains the propagators and lower-genus correlation functions. 

This yields, again up to holomorphic ambiguity at every final integration step,  
$$\eqalign{
F^{(2)}=& 
-{1\over 8} S_2^2F^{(0)}_{, 4} + 
{1\over 2}  S_2F^{(1)}_{, 2} + 
{5\over 24} S_2^3(F^{(0)}_{, 3})^2  - 
{1\over 2} S_2^2F^{(1)}_{, 1}F^{(0)}_{, 3} + 
{1\over 2} S_2(F^{(1)}_{, 1})^2+f^{(2)}\cr 
&\cr
F^{(3)}=&
S_2F^{(2)}_{, 1}F^{(1)}_{, 1}
-{1\over 2}S_2^2F^{(2)}_{, 1}F^{(0)}_{, 3} 
+ {1\over 2} S_2F^{(2)}_{, 2}
+ {1\over 6}S_2^3(F^{(1)}_{, 1})^3 F^{(0)}_{, 3}
- {1\over 2}S_2^2F^{(1)}_{, 2}(F^{(1)}_{, 1})^2 \cr &
- {1\over 2}S_2^4(F^{(1)}_{, 1})^2(F^{(0)}_{,3})^2 
+ {1\over 4}S_2^3(F^{(1)}_{, 1})^2F^{(0)}_{, 4}
+ S_2^3F^{(1)}_{, 2}F^{(1)}_{, 1}F^{(0)}_{, 3}
- {1\over 2}S_2^2F^{(1)}_{, 3}F^{(1)}_{, 1} \cr &
- {1\over 4}S_2^2(F^{(1)}_{, 2})^2
+ {5\over 8}S_2^5F^{(1)}_{, 1}(F^{(0)}_{, 3})^3
- {2\over 3}S_2^4F^{(1)}_{, 1}F^{(0)}_{, 4}F^{(0)}_{, 3}
- {5\over 8}S_2^4F^{(1)}_{, 2}(F^{(0)}_{, 3})^2 \cr &
+{1\over 4}S_2^3F^{(1)}_{, 2}F^{(0)}_{, 4}
+{5\over 12} S_2^3F^{(1)}_{, 3}F^{(0)}_{, 3}
+ {1\over 8}S_2^3F^{(0)}_{, 5}F^{(1)}_{, 1} 
- {1\over 8} S_2^2F^{(1)}_{, 4} 
- {7\over 48} S_2^4F^{(0)}_{, 5}F^{(0)}_{, 3} \cr &
+ {25 \over 48}S_2^5F^{(0)}_{, 4}(F^{(0)}_{, 3})^2 
-{5\over 16}S_2^6 (F^{(0)}_{, 3})^4   
- {1\over 12} S_2^4(F^{(0)}_{, 4})^2
+{1\over 48} S_2^3F^{(0)}_{, 6} 
+f^{(3)}\ ,}$$
where $S_2 \equiv S^{zz}$ and $F^{(g)}_{,n}\equiv (D_z)^n F^{(g)}.$

Notice that in the local case the $\chi$ drops out, as expected. 
These terms represent the various degenerations of the genus $g$ Riemann
surface, and it is useful to keep track of the calculation by interpreting 
them as Feynman rules for the auxiliary finite quantum system. Clearly the
symmetry factors are of the utmost importance to the calculation. 
We exhibit these and all 41 diagrams which contribute in the local limit to 
$F^{(3)}$ 
in 
Fig. 3.

\subsec{The local $\IP^2$ case}

${\bf P}^2$ is torically defined by the integral  lattice polyhedron 
$\Sigma={\rm conv}\{(-1,-1),
(1,0),(0,1)\}$. The non-compact A-model geometry\foot{A simple 
global Calabi-Yau, in which this geometry is embedded, is e.g. defined by 
the vanishing locus of a degree 18 polynomial in ${\bf P}^4(1,1,1,6,9)$. Using 
mirror symmetry 
some Gromov-Witten invariants for this space have been calculated \globalcalabiyau 
.}
${\cal O}(-3)\rightarrow {\bf P}^2$  is torically described by a fan spanned by 
$\{(1,-1,-1),(1,1,0),(1,0,1)\}$. By the mirror construction of \batyrev\ 
applied to the local case \kkv , the mirror B-model geometry follows then 
from the periods of a meromorphic differential $\lambda$ on the Riemann 
surface 
described by the vanishing of the $(1,1)$-shifted Newton polynomial associated to 
the
dual polyhedron 
$\Sigma^*={\rm conv}\{(-1,-1),(2,-1),(-1,2)\}$. In this case,
we simply have a cubic 
$P=x_1^3+x_2^3+x_3^3-3\psi x_1x_2x_3=0$ in ${\bf P}^2={\bf P}_{\Sigma^*}$.   

Differential equations for the periods of $\lambda$ in the  mirror geometry  
follow from the ${\bf C}^*$ scaling actions on 
the local A-model geometry. As described in \kkv\ these translate directly 
into Picard-Fuchs differential operators, which in the ${\bf P}^2$
case gives\foot{Here we have set 
$x= 27 z = 1/\psi^3 = 27(-1/a_0^3)$
(in comparison with \ckyz) to bring the singularities to 
$0,1,\infty$.}
\eqn\diffop{{\cal L}=
\theta^3-x \prod_{i+1}^3(\theta-a_i+1)=\tilde {\cal L} \theta , 
\quad a_1={1\over 3},\ \ a_2={2\over 3},\ \ a_3=1, }
where $\theta = x {\dd\over \dd x}.$
This  is the defining equation for the Meijer G-function  
{\ninepoint{$G^{2,2}_{3,3}\left(\!\! -x\! \left| \matrix{ \!\! 
{a_1}\!\!\!&\!\!\!{a_2}\!\!\!&
\!\!\!a_3\!\!\!\cr  \!\!0 \!\!\!& \!\!\!0\!\!\!& \!\!\!0\!\!\!} \right.\right)$}}, 
which we denote $G(a_1,a_2,a_3;x)$ 
for short \bateman .\foot{This
function is the  logarithmic integral 
$ {\Gamma(a_1) \Gamma(a_2)\over \Gamma(a_3)} \int {\dd x\over x} _2F_1$ 
of the hypergeometric function $_2F_1(a_1,a_2;a_3,x)$ solving $\tilde {\cal L}$. 
Solutions
$\tilde \omega_i$
of  $\tilde {\cal L}$ are related to the periods of $P=0,$ 
$\omega_i=\int_{\gamma_i} \Omega$ with 
$\Omega={\dd x\over y},$ by $\omega_i={\tilde{\omega}_i\over \psi}$. To check this 
one may use the 
Weierstrass form of the $\Gamma(3)$ curve $P=0$ $y^2=4 x^3-x g_2- g_3$  with 
$g_2=3 \psi (8+\psi^3)$, $g_3= 8+20 \psi^3-\psi^6$.
The general theory of Picard-Fuchs equations for elliptic curves then gives
for $\omega_i$ the differential
equation  $(\psi^3-1)\omega''+ 3 \psi^2\omega'+\psi \omega=0$, the same
as that which follows from $\tilde {\cal L}(\psi \omega)=0$.}

To find the solutions to \diffop~which correspond to actual integrals
of $\lambda$ over cycles, we 
relate them to the periods over the vanishing cycles. 
Both periods (apart from the trivial residue) 
are fixed up to an overall normalization by the Lefschetz theorem on vanishing 
cycles:

\noindent $\bullet$
$t_d={\partial {\cal F}\over\partial t}$ from the logarithmic solution
$t_d \log(1-27 z)+hol.$ at 
$z={1\over 27}$.

\noindent $\bullet$
$t=\sigma_0 \log(z)+\sigma_2$ from the double-logarithmic
solution at the large complex structure limit 
$z=0$: $\log(z) (\sigma_0 \log(z)+2 \sigma_1)$, where $\sigma_0$ has no constant 
term. 

It turns out that $t(x)=i {\Gamma(a_3)\over 2\pi \Gamma(a_1) \Gamma(a_2)} 
G(a_1,a_2,a_3;x)$ is precisely the mirror map\foot{The implicit choice of the 
constant term 
$\log(27)$ fits the definition of the mirror map,
a fact which turns out to be right for all local mirror
symmetry systems whose Picard-Fuchs equation is Meijer's equation.} and $t_d$  
follows by analytic 
continuation of the vanishing period at $x=1$ to produce at $z=0$ the basis of 
periods
\eqn\basis{\left(\matrix{{\partial {\cal F}\over \partial t_i}\cr
                           1\cr
                           t_i}\right)=\left(\matrix{-{d_{ijk}\over 2}t_i 
t_k+a_{ik}t_k+{1\over 24} c_2 J_i+O(q)\cr
                                              1\cr
                                              t_i}\right)}
with $d=-{1\over 3},$ $a=-{1\over 6}$ and $c_2J=-2$ as expected from the local 
limit. 
Here we have chosen a normalization\foot{If we simply take the limit of the 3-fold 
periods, 
the normalization of $t_d$ differs from this choice by $-3$.} 
of $t_d$ such that the derivative $\partial_t^3 {\cal F}$ of the period which 
stays 
finite in the local limit, ${\partial {\cal F}\over \partial t}=-{1\over 
3}(\partial_{t_E}-3 
\partial_{t_B}),$ reproduces $C_{ttt}$. This choice is also natural in the sense 
that the value
of $c_2J=-2$ reproduces the leading behavior of $F^{(1)}_{top}={c_2J \over 
12}t+\ldots$. 

With this normalization and $u=1-x$, one has near $u=0$:  
\eqn\con{\eqalign{
2 \pi i\, t_d&=A\, u+ \;{\rm higher\; order},\cr
2 \pi  i\,  t&= 2 \pi i\ t(1)+  B\, u + C\, u\, \log u+ \;{\rm higher\; order},}}
with $A=3^{-3/2}\, i$, $B=i {{(\log(27)+1) \sqrt{3}}\over {(2 \pi) }}$, 
$C={\sqrt{3}\over (2 \pi)}$.

At this point, we can use the three-point functions to solve the ${\bf P}^2$
topological partition functions
at higher-genus as we have outlined in this section.
We will use this analysis to investigate the
asymptotic growth of the $n^0_d$ in the next section.
The numerical invariants extracted from these higher-genus calculations
were listed in Fig. 2.  These B-model calculations yield results
for very high degree, though adding genera requires fixing the holomorphic
ambiguity.

Solving the B-model recursion relations gives the
topological partition functions.  As with knowing the
prepotential in (genus zero) mirror symmetry, this is
not enough to make enumerative predictions.  First one
needs to find the K\"ahler parameters $t_i$ (the ``mirror map'').
In the present case, the mirror map states that the K\"ahler
parameter $t$ is equal to the logarithmic solution of the
Picard-Fuchs equation \diffop\
of the local geometry (no ratio is needed
since the holomorphic solution is constant).
Then one must organize the prepotential
according to \gcovers, as outlined in section
2.4, to account for multiple-cover contributions.
The result is an expansion in terms of integer coefficients
$n^g_{\{d_i\}}.$

Of course, this process demands that the ambiguity (the holomorphic
section $f^{(g)}$) at each step in the recursion be fixed.  
There are $2g-1$ unknowns in this function, which we fix
by matching Gromov-Witten invariants for the
first $2g-1$ degrees at genus $g.$

There is another way to fix one of these coefficients.
Ghoshal and Vafa  argue in \ghovaf\ that the
leading coefficient of the ambiguity $f^{(g)}$ at the conifold 
singularity is determined by the free energy of the $c=1$ string
evaluated at the self-dual radius. Recall
that $f^{(g)}$ has the form $\sum_{k=0}^{2g-2} A^{(g)}_k u^{-k},$
where the conifold is at $u=0.$\foot{Ghoshal
and Vafa employ a double-scaling limit $\lambda \rightarrow
0$ and $u \rightarrow 0$ with $\lambda/u$ fixed,
where $\lambda$ is the string coupling.  This isolates
the leading behavior of $F^{(g)}$ from the subleading
terms in $F^{(g+h)}.$}  It turns out that
the leading behavior of $F^{(g)}$ is determined by $A^{(g)}_{2g-2}$
alone, as the other terms both in the anomaly and
from the recursion have subleading contributions.
Ghoshal and Vafa's identification then gives
$$A^{(g)}_{2g-2} = {B_{2g}\over 2g(2g-2)},$$
where the $B_{2g}$ are Bernoulli numbers.
We should comment that as $f^{(g)}$ is a section
of the vacuum line bundle ${\cal L},$ this equation
only makes sense in a certain gauge.  One can use the $g=0$ contribution 
to fix the gauge, then determine the higher $A^{(g)}_{2g-2}.$
Note that the independent determination of the ambiguity via
localization calculations gives a corroboration of this formula
at $g=2,3.$  In Fig. 2, the entries $n^4_{d\geq 6}$ rely on this
procedure.

This is how Fig. 2 was derived from the B-model.

\newsec{Some Enumerative Issues}

The holomorphic curves that these numbers count are not
simply worldsheet instantons.  As shown by Vafa and Gopakumar \vafagop,
they represent D-branes in
type-II or M-theory compactifications. 
In this section, we will analyze the growth of these
invariants at genus zero, then make some remarks regarding
a proper mathematical interpretation of the integers.

\subsec{Asymptotic growth of states}

Let us return now to the solutions \con\
of the ${\bf P}^2$ period equations.

The singularity  at $x=1$ limits the radius of convergence of
the instanton expansion and in particular 
the exponential growth of $|n_d|$ must be  $e^{2\pi t_2(1)},$
where $t_2 = {\rm Im}(t).$  For an 
interpretation of the $n_d \equiv n^0_d$ as BPS counts for some physical system, 
the logarithm of the multiplicity of the BPS states is the entropy in the large 
$d$ limit 
and an important critical exponent of the theory.  The value 
\eqn\asym{t(1)= {i\over {2\pi \Gamma({1\over 3}) \Gamma({2\over 3})}} 
G\left({1\over 3},{2\over 3},1;1\right)\sim -{1\over 2} +i\ 0.462757788\ldots ,} 
is therefore of particular interest. The real part ${1\over 2}$ means that the 
$n_d$ 
come with alternating sign. 
Fig. 4 shows how the logarithmic slope of $|n_d|$ approaches $t_2(1)=Im(t(1))$.

\vskip 0.5cm
{\baselineskip=12pt \sl
\goodbreak\midinsert
\centerline{{\epsfxsize 2.5 truein\epsfbox{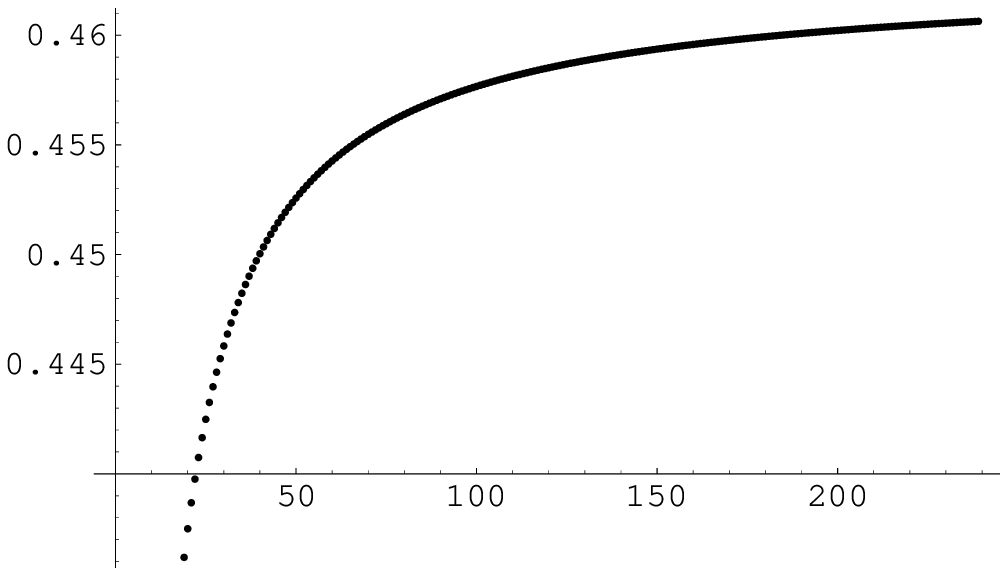}} \ \ \ \ \  { \epsfxsize 2.2 
truein\epsfbox{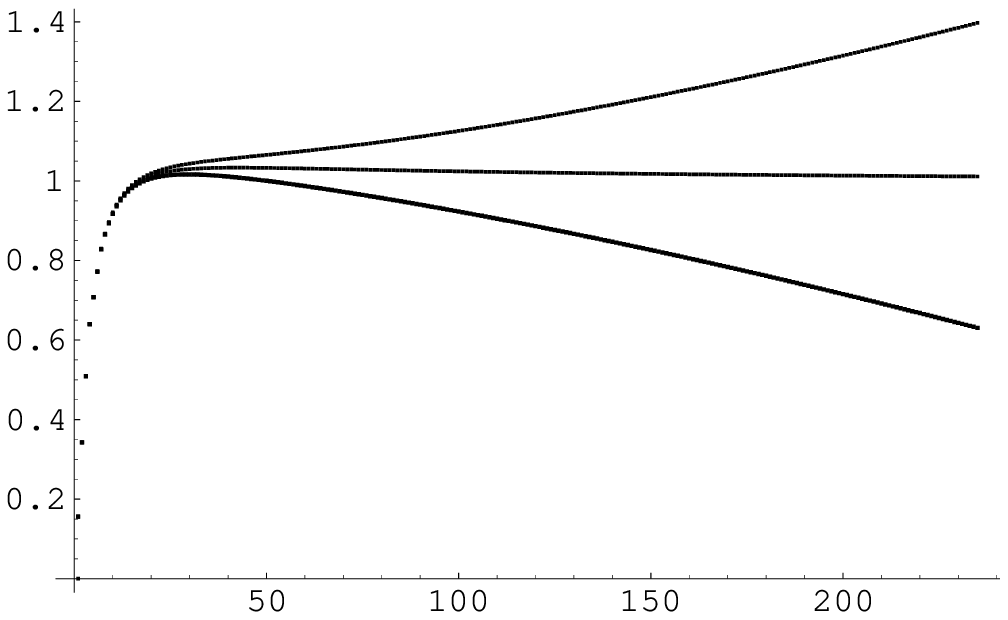}}}
\leftskip 1pc\rightskip 1pc \vskip0.3cm
\noindent{\ninepoint  \baselineskip=8pt  
{\bf Fig. 4:} Left: the slope of ${\log(|n_d|)\over 2 \pi}$ for $d=1,\ldots 240$ 
approaches 
$t_2(1)$ for high $d$. Right: Sensitivity of the third Richardson transform of 
$n_d/n_d^{asym}$ on $t_2(1)$. 
The curve approaching $1$ is for correct $t_2(1)$ value \asym~ the lower/upper 
curve is for $t_2(1)\pm\varepsilon$
with $\varepsilon=2\times 10^{-6}$.}
\endinsert}

More precisely, one can also determine as in \cdgp \ the first subleading 
logarithms by comparing 
the ansatz for $n_d^{asym}\sim N d^{\rho} \log (d)^\sigma e^{2 \pi d t_2(1)}$ in 
the asymptotic 
expansion with $C_{uuu}$.
The Yukawa coupling in the $x$ variable 
\eqn\yukpII{F^{(0)}_{xxx}=-{1\over 3 (2 \pi i)^3} {1\over x^3(1-x)}} 
is  normalized so that that $F^{(0)}_{ttt}=\left({\dd z \over \dd t}\right)^3 
F^{(0)}_{zzz}=-{1\over 3} + \sum_{d=1} {d^3 n_d q^d\over 1-q^d}$ reproduces the 
genus 0 
instantons 
as in Fig. 2. This and the knowledge of the $u \log u$ coefficient in the $t(u)$ 
expansion     
yields 
\eqn\assym{n_d^{asym} \sim {-\int J^3 \over C^2} {e^{-2 \pi i d t(1)}\over d^3 
\log (d)^2 } \ . }
Instead of just comparing ${n_d/ n_d^{asym}}$ one can suppress further subleading
logarithms by considering higher order Richardson transforms \cdgp , which also 
approach one. 
This gives a far more sensitive check on the
$t_2(1)$ slope from the instanton numbers --
see Fig. 4. 

The formula \asym\ describes the asymptotic of other one-modulus local mirror 
symmetry 
systems with Picard-Fuchs equations
given by Meijer's equation with solution $G(a_1,a_2,a_3;x)$. 
These cases were discussed in \kmv\lmw\ckyz . If we define
 $b \equiv b(\vec a) \equiv 2 \Psi(a_3)-\Psi(a_1)-\Psi(a_2)$
(here $\Psi(x) = d (\log \Gamma(x))/dx$ is the
digamma function), then we have:

\noindent $\bullet$ $({1\over 2}, {1\over 2},1)$: describes the diagonal direction 
in ${\bf P}^1\times 
{\bf P}^1$ with
$x=e^b z$ \ckyz\ as well as a direction in the $E_5$ del Pezzo,
with $x=-e^b z$.

\noindent $\bullet$ $({1\over 3}, {2\over 3},1)$: 
describes $\IP^2$ as discussed, with $x = e^b z$, 
as well  
as a direction in the $E_6$ del Pezzo, with $x=-e^b z$. 

\noindent $\bullet$ $({1\over 4}, {3\over 4},1)$: describes a direction in the 
$E_7$ del Pezzo, with $x= - e^b z$ .

\noindent $\bullet$ $({1\over 6}, {5\over 6},1)$: describes a direction in the 
$E_8$ del Pezzo, with $x= -e^b z $.
 
With $B(\vec a)\equiv i{b(\vec a)+1\over 2 \pi}$ we summarize some of their
relevant properties, which 
follow to a large extent from Meijer's fundamental system of solutions.

{\vbox{\ninepoint{
$$
\vbox{\offinterlineskip\tabskip=0pt
\halign{\strut\vrule#
&\hfil~$#$
&\vrule#
&~\hfil ~$#$~
&\hfil ~$#$~
&\hfil $#$~
&\hfil $#$~
&\hfil $#$~
&\hfil $#$~
&\hfil $#$~
&\hfil $#$~
&\vrule#\cr
\noalign{\hrule}
&                && \int J^3& \int c_2J &  a        &  A              &  B    &    
    C    & t(1)        & \Gamma &\cr
\noalign{\hrule}
&P^1\times P^1   &&  -1     &    2      & 0         &-{i\over 2}      &2B(\vec a)& 
{1\over \pi}
&.371226\, i&  \Gamma^0(4)      &\cr
& E_5            &&   -4    &   -^*4    & -{4\over2}        & -\sqrt{4}i           
  &\sqrt{4}B(\vec a)& {1\over \pi}
&-{1\over 2}+.371226\,i&\Gamma_0(4)&\cr
\noalign{\hrule}
& P^2            &&-{1\over3}&  -2      &-{1\over 6}& {i\over 
3^{3/2}}&\sqrt{3}B(\vec a)      & {\sqrt{3}\over 2 \pi}
&-{1\over 2}+.462757\, i& \tilde \Gamma^0(3)&\cr
& E_6            && -3      &   -^*6    &-{3\over 2}& i\sqrt{3}        
&\sqrt{3}B(\vec a)     & {\sqrt{3}\over 2 \pi}     
&-{1\over 2}+.462757\, i&  \Gamma_0(3)&\cr
\noalign{\hrule}
& E_7            && -2      &   -8      &-{2\over 2}         &-i\sqrt{2}        
&\sqrt{2}B(\vec a)     & {\sqrt{2}\over 2 \pi}     
&-{1\over 2}+.610262\, i &\Gamma_0(2)&\cr
& E_8            && -1      &   -10     &-{1\over 2}& -i               &B(\vec a)     
& {1\over 2 \pi}    
&-{1\over 2}+.928067\, i  &\Gamma(1)        &\cr
\noalign{\hrule}}
\hrule}$$
\vskip-7pt
\vskip7pt}}

The exact value of the critical exponent is $G(a_1,a_2,a_3;1)$ and the asymptotic 
expansion follows from \assym .
Note that for all del Pezzo surfaces, $-{\int J_3\over C^2}= (2\pi)^2$ in good 
agreement with the instanton
numbers.\foot{This can be seen from the third Richardson transform.
The asymptotic formula for the $E_8$ del Pezzo 
given in \mnw~is not quite correct. 
The monotonically increasing function $n_d/n^{asym}_d$ would cross $1$ at about
$d=100.$} If there is a star on $\int J c_2,$ we find an additional constant shift
by one in the double-logarithmic 
solution \basis . 
The monodromies are generated by the two  shift operators at $x=0$ and $x=1$,
which follow from the 
constants listed in the table. 
   
There is an important difference between the $({1\over 2},{1\over 2},1)$ system 
and the other cases.  It follows from the Riemann symbol that
the former has a logarithmic solution at ${1\over z} = 0$, while the others have, 
as in the 
one-modulus 
compact three-folds \kt, power series solutions at this point
signaling a 
conformal
field theory.  In the compact case these can be
identified as Gepner-type conformal
field theories:  tensor products of minimal $N=2$ models.

The analytic continuation to ${1\over z}=0$ is particularly
simple due to a Barnes integral representation -- see \cdgp\kt\agm\ for similar 
applications.
As in \agm~it is easy to see that, given the above definition 
of $t$ at $z=0$, which corresponds to a choice of the $B$-field, 
one gets $t={1\over 2}$ at $1/z=0$, if $z=e^bx$ and $t=0$ and 
at $1/z=0$ if $z=-e^b x$. In particular, the $\IP^2$ example has a 
conformal point at $1/z=0$, which is the ${\bf C}^3/{\bf Z}_3$ non-compact 
orbifold.

\vskip 0.5cm
{\baselineskip=12pt \sl
\goodbreak\midinsert
\centerline{\epsfxsize 3 truein\epsfbox{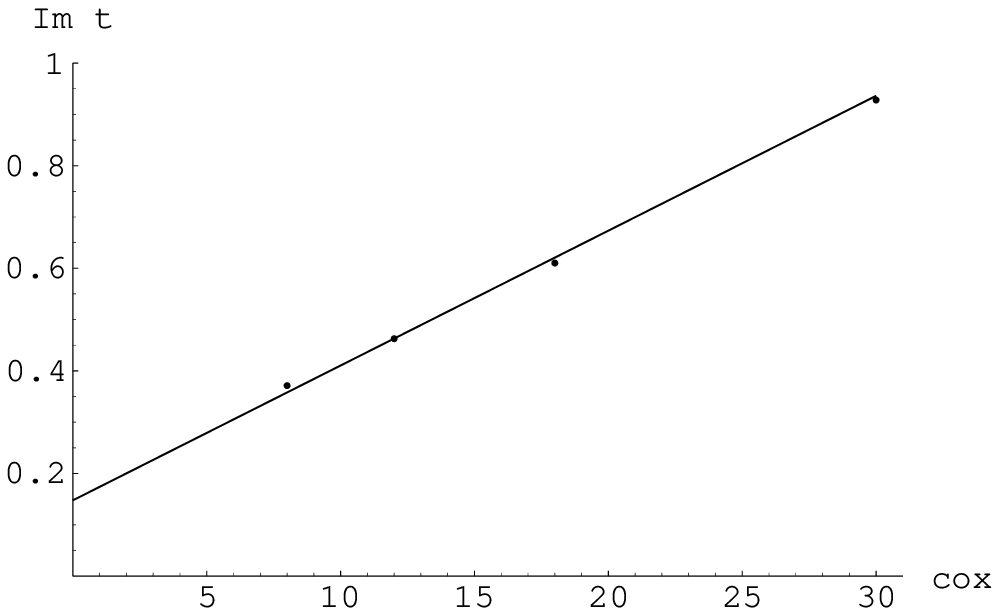}}
\leftskip 1pc\rightskip 1pc \vskip0.3cm
\noindent{\ninepoint  \baselineskip=8pt  
{\bf Fig. 5:} Near linear dependence of the
growth rate of instantons with the Euler (Coxeter) number.}
\endinsert}

Clearly it would be extremely interesting to find the physical
system whose entropy 
of
BPS states approaches in some
limit the critical exponents which we have provided here. 
It is noteworthy that the coefficients $t_2(1)$ for the del Pezzo cases grow
linearly (to a close approximation)
with the Euler number assigned to the corresponding local Calabi-Yau,
i.e. twice the Coxeter number of the associated group.
If, as for K3, there is a bosonic
string-like oscillator algebra on the moduli spaces, this
would be the expected behavior.  
For example, in the local ${\bf P}^1\times{\bf P}^1$ case,
an oscillator-type partition function for
some of the local invariants was found
by Nekrasov and Kol \nek, though the sense in which
the coefficients were Euler characteristics of a moduli
space remains unclear.
This brings us to our next question.

\subsec{What is the moduli space?}

{}From the work of Gopakumar and Vafa II \vafagop, we learn that
the integers arising from a proper organization of the
topological string partition functions simply encode
the number and properties of BPS states in the specified
charge (homology class).  Note that the BPS charges are specified by
a homology class, not a genus.  In fact, different genera
are part of a connected moduli space of BPS states of
specified charge.

The integers describe the representation content of
the BPS states under an additional conjectural Lefschetz-type
$SU(2)$ action on the cohomology of this moduli space.  As is
familiar from BPS counting on K3 \bsv \yz, this moduli space
is naively comprised of pairs of holomorphic curves
together with flat $U(1)$ connections.  This thinking may be
too naive.

What is the proper definition of the moduli space?
Experience is teaching us that D-branes can be thought of
as objects in the bounded derived category of coherent sheaves.
Indeed, this category is important for Kontsevich's homological
mirror conjecture \kontsevich.
The conjecture is that the derived category of coherent sheaves
(D-branes/instantons in IIA/B) is equivalent to a category of Fukaya whose objects
are Lagrangian (perhaps special-Lagrangian) objects (D-branes/instantons in
IIB/A).  
This conjecture satisfies the correspondence principle. 
For example, it captures the idea of
constructing the mirror Calabi-Yau through special-Lagrangian tori \syz\
as the equivalence of the moduli spaces of two distinguished objects
(the structure sheaf of a point versus the distinguished toroidal
Lagrangian submanifold).  It also includes work of Vafa \vafaslag\ by
interpreting the three-point functions of massless excitations
above a D-brane state as structure constants in the composition
of morphisms -- though Vafa has taken this further by identifying
special coordinates through which to make identifications across
mirror theories.  Finally, ordinary mirror symmetry would be
recovered by looking at similar considerations by treating the
Calabi-Yau three-fold $M$ as a (diagonal) D-brane in $M\times M.$ 
  
There is further evidence for the derived category.
Hilbert schemes of points are a kind of intermediate
step between moduli spaces of vector bundles and full-out derived
moduli spaces, and they appear prominently in D-brane counts. 
Evidence also
comes from requiring the Fourier-Mukai transform to act as a symmetry
of objects for K3 compactifications \aspdon.  As well, D-brane
charges naturally live in K-theory \minmoor \kthy.
For ${\bf P}^2$ these moduli spaces of objects are not well known, though some
Poincar\'e polynomials of moduli spaces of 
points (sheaves supported on points or schemes of finite length)
have been computed, as have some moduli spaces of higher-rank
bundles.  Heisenberg algebras have been found to act on cohomology,
as for K3, which we take as loose evidence that these are the right spaces
to study \vertalgs.  In addition, the exponential growth of the
local invariants (with exponent
depending on the ``non-compact Euler number'') is in line with
(super-)oscillator partition functions.

Let us now think of sheaves as one-term complexes,
so as objects in a derived category.  Then the moduli
spaces of objects in this category should  be the proper general
framework for these questions.  Moduli spaces of derived objects
in general do not exist due to obstructions.  For this reason,
derived moduli spaces \foot{``Derived quot schemes,'' in fact, which retain
all obstruction data in the form of a complex.  In short, dg-schemes
do not have tangent {\sl spaces,} they have tangent
{\sl complexes.}} \quot\ may be the technology
necessary to study the higher-genus integer invariants in the general case.
Euler characters for such ``spaces'' can be defined.
One can hope that this formulation of the question will lead to
insight regarding the higher-genus problem for the quintic, which
is not amenable to localization calculations.

Thinking of derived moduli spaces rather than moduli of sheaves
is more in line with the spirit of Kontsevich's mirror symmetry
conjecture \kontsevich.  His motivation for
introducing an $A_\infty$ structure came in part from
trying to define an extended moduli space of Calabi-Yau manifold
as the moduli space of the structure sheaf of the diagonal $M$ in
$M\times M$ by deforming the
associative algebra of functions as a
homotopy associative differential graded ($A_\infty$) algebra.
The relation to the algebra to BRST cohomology classes
on the conformal field theory of the D-brane, and of the
grading to ghost number would be interesting to explore.

In fact, it is likely
that we must treat our sheaves as sheaves in
the total space of the canonical bundle of ${\bf P}^2.$
This would introduce further complexity.

\subsec{Counting singular curves}

For many applications, the differences between D-brane as
cycle with bundle, as sheaf, or as derived object are unimportant.
However, we may not be so lucky here.
The argument in the
second paper of \vafagop\ gives a heuristic derivation of
$n^0_3({\bf P}^2) = 27.$  This should be equal to
$\chi({\cal M}_3),$ the Euler
characteristic of the D-brane moduli space ${\cal M}_3$ of curves
in the class $3H,$ where $H$ is the hyperplane class generating
$H_2({\bf P}^2).$  The $n^g_d$ for $g\neq 0$ represent the
different characters of the additional fiber-Lefschetz $SU(2)$ action on
cohomology.
$n^0_3({\bf P}^2)$ is obtained by noting that
the Jacobian of a smooth degree three curve can be identified
with the curve itself (which has genus one).  Therefore the choice
of curve plus point on Jacobian is the same as the choice of
curve plus point on curve.  If we choose the point first (a ${\bf P}^2$
worth of such choices) and ask how many
degree three curves pass through the point (a ${\bf P}^8$), we
see ${\cal M}_3$ to be a ${\bf P}^8$ fibration over ${\bf P}^2,$
with Euler characteristic $9\times 3 = 27.$

Following $\yz,$ one might attempt to count the Euler characteristic by
noting that ${\cal M}_3$ must be a fibration over the moduli space
of degree three curves in ${\bf P}^2,$ with fiber equal to the
Jacobian of a given curve.  As Jacobians of smooth curves have
trivial Euler characteristic, we can localize our counting to the
singular (more specifically, the rational curves).  There is a
moduli space of such curves (for K3 this space was zero dimensional),
each of which should contribute some fixed amount determined by
the singularity type.  However, among the degree three curves we
necessarily have reducible (e.g., $(X+Y)(X^2 + Y^2 + Z^2) = 0 \subset {\bf P}^2$)
and even unreduced (e.g., $(X+Y)^3 = 0 \subset {\bf P}^2$) curves.
Here is where the sheaf interpretation is necessary.
Over the unreduced curves, which can be thought of as multiply-wrapped
branes, we should have rank $k$ bundles (if $k$ is the number of
wrappings).  These sheaves then have the same characteristic
numbers as the non-singular curves.  We are still unsure of how
to perform the counting (e.g., there is a discrete infinity of
higher-rank bundles over ${\bf P}^1$ with fixed degree).  This 
interpretation is consistent with the moduli spaces of objects
in the elliptic curve.

We can now hope to state Gopakumar and Vafa's
approach as a conjecture that the Euler characteristics
of the appropriate moduli dg-stack of derived objects on
${\bf P}^2$ of characteristic numbers
$(r=0,c_1=d,c_2=d^2)$ have Euler characteristic equal to $n^0_d.$
This conjecture might have to be tailored to account for the
fact that the objects properly live on $K_{{\bf P}^2}$ with support
in ${\bf P}^2.$  Whether we need the full weight of dg-stacks
(which have tangent spaces which include all obstructions to
deformations) to check this, remains to be seen.
For the general situation, it is likely that we would.

These spaces may give some insight into the multiple
cover formulas of Vafa and Gopakumar, as well.
One also must find an additional fiber-Lefschetz $SU(2)$ action on the
derived moduli spaces to agree with Vafa and Gopakumar's interpretation for 
$n^{g\ge 1}_d$.
Exciting work lies ahead.

\centerline{\bf Acknowledgements}

We are grateful to S. Katz, R. Pandharipande, T. Graber and C. Vafa 
for illuminating conversations.  Also we would like to thank 
C. Faber for sending us his Maple program for the intersection 
calculation on $\overline{{\cal M}}_{g,n}$. 
The work of A. Klemm is supported in part by 
a DFG Heisenberg fellowship and NSF Math/Phys DMS-9627351.

\listrefs

\end